\documentclass[preprint,authoryear,12pt]{elsarticle}

\usepackage{color}
\usepackage{amssymb}
\usepackage{graphicx}

\usepackage[
a4paper=true,%
breaklinks=true,%
colorlinks=true,%
pdfauthor={First Author et al.},%
pdftitle={Template for manuscripts in Advances in Space Research}%
]{hyperref}

\journal{Advances in Space Research}

\newcommand{\be}{\begin{equation}}
\newcommand{\ee}{\end{equation}}

\begin{document}
\begin{frontmatter}

\title{Cosmic Ray transport in the Galaxy: a Review}
\author{Elena Amato\corref{cor}}
\address{INAF - Osservatorio Astrofisico di Arcetri, Largo E. Fermi, 5 - 50125 - Firenze, Italy\\
Dipartimento di Fisica e Astronomia, Universit\`a degli Studi di Firenze, Via Sansone, 1 - 50019 - Sesto Fiorentino (FI), Italy}
\cortext[cor]{Corresponding author}
\ead{amato@arcetri.astro.it}

\author{Pasquale Blasi}
\address{INAF - Osservatorio Astrofisico di Arcetri, Largo E. Fermi, 5 - 50125 - Firenze, Italy\\
Gran Sasso Science Institute, Viale F. Crispi, 7 - 67100 - L'Aquila, Italy}
\ead{blasi@arcetri.astro.it}

\begin{abstract}
The physics of energetic particle propagation in magnetized environments plays a crucial role in both the processes of acceleration and transport of cosmic rays. Recent theoretical developments in the field of cosmic ray research have been mainly in the direction of exploring non-linear aspects of the processes in which these particles  are involved, namely the action of cosmic rays on the environment in which the transport and/or acceleration take place. When cosmic rays propagate outside of the acceleration region, such action is mainly in two forms: 1) they generate hydromagnetic waves, through streaming instabilities, leading to a dependence of the scattering properties of the medium on the spectrum and spatial distribution of the energetic particles, and 2) they exert a dynamical action on the plasma, which may cause the launching of cosmic ray driven Galactic winds. In this article we discuss these and other recent developments and how they compare with the bulk of new observations on the spectra of primary nuclei (mainly H and He) and secondary to primary ratios, such as the B/C ratio and the $\bar p/p$ ratio, and the positrons ratio $e^{+}/(e^{-}+e^{+})$. We also comment on some radically new models of the origin of CRs, in which the physical meaning of the secondary to primary ratios is not the same as in the standard model.
\end{abstract}
\begin{keyword}
cosmic rays; ISM; diffusion; MHD.
\end{keyword}
\end{frontmatter}

\parindent=0.5 cm

\section{Introduction}
Building a satisfactory theory for the origin of cosmic rays (CRs hereafter) is a task that consists of two fundamental parts that combined together may provide a consistent picture of what we observe: a model of particle acceleration in a class (or more than one class) of sources and a model for transport of these particles from their sources to Earth. The latter part is probably the one that is attracting most attention at present, and this is mostly due to some recent observations that might be challenging to explain in the context of the standard paradigm of cosmic ray transport through the Galaxy. 

The recent measurements of experiments such as PAMELA, CREAM and AMS-02, have shown that what appeared to be a rather featureless, proton dominated, power-law spectrum from about 10 GeV to the {\it knee} energy of about {\it few} $\times 10^{15}$ eV is actually rich in structure both in terms of spectral properties and composition. First, the spectrum of protons and helium nuclei (and possibly also of all nuclear species) seems to flatten in the energy range 200-300 GeV/n \citep{2011Sci...332...69A,Ahn10,2015PhRvL.114q1103A}. Second, the slopes of these species are slightly different from each other. This latter effect leads, among other things, to the dominance of the He flux over that of H nuclei above 100 TeV. The appearance of breaks in otherwise power-law trends typically points to the existence of some energy scale that may suggest the action of a new physical phenomenon worth being investigated.

Even more surprising are the recent findings concerning the anti-matter component of CRs: an unexpected rise with energy in the positron to electron ratio has been detected by both the PAMELA \citep{2009Natur.458..607A} and AMS-02 experiment \citep{2013PhRvL.110n1102A}. In the standard scenario of CR transport, positrons are only secondary products of inelastic pp collisions, which implies that the positron ratio should decrease with energy, hence the discovery of a rise in this quantity led to many speculations, ranging from anomalous transport to new sources of positrons, either due to annihilation/decay of dark matter particles or astrophysical phenomena (see \cite{2012APh....39....2S} for a comprehensive review). The recent measurements of the ratio of the anti-proton to proton fluxes  \citep{AMS02pbar} added to the puzzle: this measurement suggests that this ratio does not drop with energy as fast as naively expected based on the standard model, although this conclusion is not universally accepted, given the uncertainties in both the cross section of ${\bar p}$ production and in CR transport. These findings have stimulated a blossoming of possible scenarios that may explain observations, even questioning the very bases of the standard model of CR transport in our Galaxy.

In the last decade or so the research on CR propagation has benefited of a considerable increase and improvement in the resources available to treat the problem numerically: GALPROP \cite[]{1998ApJ...509..212S} has been joined by DRAGON \cite[]{2008JCAP...10..018E}, PICARD \cite[]{2014APh....55...37K} and Usine \cite[]{2001ApJ...555..585M} and the description of both CR transport and of its observable consequences in terms of secondary particles and radiation has become increasingly more sophisticated. However all these numerical approaches adopt some prescriptions that have as a main motivation that of making predictions that agree with all available data (including particle spectra at the Earth, secondary-to-primary ratios and diffuse Galactic gamma-ray emission). For instance, all numerical approaches to CR propagation require breaks in the rigidity dependence of the particle diffusion coefficient or injection spectrum or both. The question that we are interested to address here concerns the existence of physical processes possibly associated with CR transport that may cause such breaks or any other feature in otherwise power law trends. 

The non-linear action of CRs on the environment manifests itself in mainly two ways: 1) through excitation of streaming instability which generates waves that in turn are able to scatter CRs, and 2) by exerting a dynamical action on the plasma, which may cause plasma motions, such as cosmic ray driven Galactic winds.

Aside from the obvious need to connect theory and observation, there is another, maybe more subtle need to look for a more solid foundation of the theory of CR transport, namely find a physical motivation for aspects that we usually embed in the boundary conditions of our mathematical formulation of the problem. For instance, the diffusive treatment of CR transport leads to the well known and widely used relations that connect injection and spectra at the Earth, only if one assumes the existence of a so-called free escape boundary at some height from the disc. Such a boundary condition might be reflecting the existence of a scale height in the level of turbulence or background magnetic field, so that the diffusion coefficient grows quickly for $|z|>H$. In this case one finds the standard relations mentioned above. However, the halo size could, just as well, naturally emerge from a theory of CR induced Galactic winds: in this case there is an energy dependent transition from diffusion dominated to advection dominated transport which plays the role of a halo boundary surface. No explicit boundary condition is required at any finite distance, and the connection between injection and equilibrium spectrum is non-standard. Distinguishing between two such possibilities may make the difference between fitting data and modelling the physics of transport.

The main aim of the paper is that of discussing some physical aspects of non-linear CR transport and illustrating some phenomenological implications, such as features in the spectrum of nuclei. Moreover, some non-linear effects might provide the basis for a better physical understanding of commonly adopted boundary conditions on the differential equations that describe CR transport. In fact we will also discuss the role that these non-linear effects might play in radically new scenarios of CR propagation. 

The article is organized as follows: in \S \ref{sec:standard} we review the standard paradigm of CR propagation and its predictions in terms of secondary to primary ratios; in \S \ref{sec:waves} we summarize the main processes involved in the growth of cosmic ray induced turbulence; in \S \ref{sec:Gal-self} we discuss the role of CR self-generated waves in the transport of these particles throughout the Galaxy, how they are expected to affect the diffusion coefficient and induce breaks in the spectra measured at Earth; in \S \ref{sec:grammage} we discuss how wave self-generation may influence the confinement of CRs in the vicinity of their sources and affect the grammage; in \S \ref{sec:Gal-wind} a scenario in which CRs are responsible for the launching of a Galactic wind is considered and its consequences on the particle spectrum discussed; finally, in \S \ref{sec:secondary} the ``anomalies'' in the secondary to primary ratios
recently highlighted by direct detection experiments are discussed; solutions that invoke non-standard propagation scenarios are reviewed in \S \ref{sec:nonstandard}, while \ref{sec:pulsars} is devoted to assessing the role of pulsars as contributors to the positron excess; we provide a summary and conclusions in \S \ref{sec:conclude}.

\section{A summary of standard predictions}
\label{sec:standard}
In the standard model for the origin and propagation of Galactic CRs, these are accelerated in the blast waves of Supernova Explosions and propagate diffusively throughout the Galaxy interacting with the interstellar gas. The SN blast wave picks up the particles to be accelerated from the Interstellar Medium (ISM), and hence the mass composition is expected to reflect that of the ISM, although injection at the shock is also known to work preferentially for high $A/Z$ nuclei \cite[]{1981JGZG...50..110E}. Electrons are also accelerated together with protons and heavier nuclei, while positrons (and antiprotons) present in CRs have long been considered to be of purely secondary origin, namely deriving from the nuclear collisions of CR hadrons with the ISM they encounter during their journey from the sources to Earth. 

Many subtle aspects of CR transport have been discussed for many years and have led the community to the adoption of propagation codes such as GALPROP \cite[]{1998ApJ...509..212S}, DRAGON \cite[]{2008JCAP...10..018E}, PICARD \cite[]{2014APh....55...37K} and Usine \cite[]{2001ApJ...555..585M}. This section is not aimed at summarizing all the details of propagation that such codes can describe, mainly thanks to a detailed description of the gas distribution in the Galaxy and the collection of accurate data on the relevant cross sections. Instead we aim at providing a description of the main physical results that can be derived from simpler descriptions of CR transport. 

The simplest implementation of CR transport that still retains the main physical aspects is one in which the effective transport only takes place in the $z$ direction perpendicular to the Galactic disc. The formalism can be easily generalized to 3D. The transport equation for CR nuclei (both primaries and secondaries) can be written as:
$$
-\frac{\partial}{\partial z} \left[D_{\alpha}(p) \frac{\partial f_{\alpha}}{\partial z}\right] + w\frac{\partial f_{\alpha}}{\partial z} -\frac{p}{3}\frac{\partial w}{\partial z}\frac{\partial f_{\alpha}}{\partial p}
+\frac{\mu v(p) \sigma_{\alpha}}{m}\delta(z) f_{\alpha} + 
\label{eq:traps}
$$
\be
\frac{1}{p^{2}} \frac{\partial}{\partial p}\left[ p^{2} \left(\frac{dp}{dt}\right)_{\alpha,ion} f_{\alpha}\right] =
\label{eq:slab}
\ee
$$
~~~~~~~~~~~~= 2 h_d q_{0,\alpha}(p) \delta(z) +\sum_{\alpha'>\alpha} \frac{\mu\, v(p) \sigma_{\alpha'\to\alpha}}{m}\delta(z) f_{\alpha'},
$$
where we defined:
\be
w(z)=(u+\bar v_{A})\Theta(z)-(u+\bar v_{A})\left[ 1 - \Theta(z) \right],
\ee
with $\Theta(z)$ the Heaviside function. Here $u$ is the velocity of a possible Galactic outflow (if none is present $u=0$). The mean Alfv\'en speed $\bar v_{A}$ is the effective Alfv\'en speed averaged upon the direction of motion of the waves: if there is an equal number of waves moving in both directions then $\bar v_{A}=0$. This is the situation in which one should expect the highest degree of second order Fermi acceleration. On the other hand, if waves only move away from the disc, as it is expected to be the case if they are self-generated, then $\bar v_{A}=v_{A}=B_{0}/\sqrt{4\pi m_{p} n_{i}}\approx 15$ km/s ($B_{0}=1\mu G$ and $n_{i}=0.02~\rm cm^{-3}$ are estimates of the magnetic field strength and gas density in the halo). In this case there is no second order Fermi acceleration. 

In Eq. (\ref{eq:slab}), $\sigma_{\alpha}$ is the spallation cross section of a nucleus of type $\alpha$, $\mu$ is a grammage parameter of the disc fixed to 2.4 mg/cm$^2$ ($\mu=m_{\rm eff} n_H 2 h_{\rm disk}$, with $m_{\rm eff}\approx 1.4 m_p$, $n_H$ the gas density in the disk and $h_d$ the disk height), and $q_{0,\alpha}(p)$ is the rate of injection per unit volume in the disk of the Galaxy. 
The total cross section for spallation and the cross sections for the individual channels of spallation of a heavier element to a lighter element ($\sigma_{\alpha'\to\alpha}$) are provided by \cite{1990PhRvC..41..566W} and \cite{2003ApJS..144..153W}. 

At high enough energies that ionization losses (fifth term in Eq. \ref{eq:slab}) are not important and advection terms, even when present, are subdominant, considering protons as primary nuclei, the equation above reduced to the well known equation:
\be
-\frac{\partial}{\partial z} \left[D_{p}(p) \frac{\partial f_{p}}{\partial z}\right] =
2 h_d q_{0,p}(p) \delta(z) 
\label{eq:slabprotons}
\ee
where the injection term is $2 h_d q_{0,p}(p) = \frac{N(p){\cal R}}{\pi R_{d}^{2}}$, $N(p)$ is the spectrum produced by each source (for instance a SNR), ${\cal R}$ is the rate per unit time of occurrence of such sources in the Galaxy and $R_{d}$ is the radius of the Galactic disc. Notice that while Eq.~\ref{eq:slabprotons} is specialized to the simplest case of a uniform injection throughout the disc, the same equation can be easily generalized to include a dependence of the injection rate on the galactocentric radius.

Eq. \ref{eq:slabprotons} is usually solved by imposing a free escape boundary condition, $f_{p}(|z|=H,p)=0$ and using the symmetry of the problem, $\frac{\partial f_{p}}{\partial z}|_{z=0^{-}}=-\frac{\partial f_{p}}{\partial z}|_{z=0^{+}}$. These assumptions lead to the well known solution: 
\be
f_{0,p}=\frac{N(p) {\cal R}}{2 \pi R_{d}^{2} H}\frac{H^{2}}{D(p)}\ .
\label{eq:standardspec}
\ee
This latter expression has a simple physical interpretation: the density of protons of any given momentum as observed at the Earth is the product of the rate of injection per unit volume on the entire volume of the Galaxy and the confinement time $H^{2}/D(p)$. For an injection $N(p)\propto p^{-\gamma}$ and a diffusion coefficient $D(p)\propto p^{\delta}$, the CR spectrum inside the disc is $f_{0,p}\propto p^{-\gamma-\delta}$. 

For primary nuclei the situation is similar but energy losses due to spallation may become important (see for instance \cite{2012JCAP...01..010B}), especially for heavier species. In the simplest case in which there is no advection and spallation occurs mainly in an infinitely thin disc, one can write a simpler version of Eq. \ref{eq:slab} as:
\be
-\frac{\partial}{\partial z} \left[D_{\alpha}\frac{\partial I_{\alpha}(E_{k})}{\partial z}\right] =
\frac{N_{\alpha}(E_{k}) {\cal R}}{\pi R_{d}^{2}} \delta(z)  - I_{\alpha}(E_{k}) n_{d} v(p) \sigma_{\alpha} 2h \delta(z),
\ee
where we assumed that the contribution of spallation of heavier primaries to the flux of a given primary nucleus is negligible. Here we also introduced the flux of nuclei of type $\alpha$, as $I_{\alpha}(E_{k})dE_{k}= p^{2} v(p) f_{\alpha}(p) dp$, so that $I_{\alpha}(E_{k}) = A p^{2}f_{\alpha}(p)$ \cite[]{2001ApJ...547..264J}, with $E_{k}$ the kinetic energy per nucleon for a nucleus of mass number $A$. The solution of this equation is easily found following the same steps as above:
\be
I_{\alpha}(E_{k}) = \frac{N_{\alpha}{\cal R}}{2 \pi R_{d}^{2} H}\frac{H^{2}}{D}\frac{1}{1+\frac{X(R)}{X_{cr,\alpha}}}.
\ee
Here we introduced the grammage $X(R)=n_{d}(h/H) v \frac{H^{2}}{D(R)}$, as a function of the rigidity $R$ and the critical grammage $X_{cr,\alpha}=m_{p}/ \sigma_{\alpha}$ for a nucleus of type $\alpha$. Since the grammage is a decreasing function of rigidity, one can expect that at high rigidities the grammage traversed by particles is small ($X(R)\ll X_{cr,\alpha}$) so that the standard scaling, also found above for protons, is recovered. On the other hand, in the opposite limit $X(R)\lesssim X_{cr,\alpha}$, the observed spectrum gets harder, namely closer to the injection spectrum.

For all nuclei, including protons, ionization losses become important at low energies. What happens to the spectrum then depends on a rather complex interplay between advection (when present), diffusion (in this energy region the diffusion coefficient is usually, but not always, assumed to be constant) and losses. 

For secondary nuclei, the situation is more complex because the whole chain of spallation reactions of heavier nuclei should be followed. However a physical understanding of what happens can be achieved by assuming that the main contribution to the flux of a given secondary product is provided by one dominant primary. The process of spallation conserves the kinetic energy per nucleon $E_{k}$, hence, repeating the calculations above but taking into account that a secondary nucleus is both produced by the parent nucleus and destroyed by spallation, one has:
\be
I_{\alpha}(E_{k}) = I_{\alpha'>\alpha}(E_{k})\frac{\frac{X(R)}{X_{cr,\alpha'\to\alpha}}}{1+\frac{X(R)}{X_{cr,\alpha}}},
\label{eq:sec}
\ee
where $X_{cr,\alpha'\to\alpha}=m_{p}/\sigma_{\alpha'\to\alpha}$ and $X_{cr,alpha}=m_{p}/\sigma_{\alpha}$. Eq. \ref{eq:sec} shows the load of physical information embedded in the so called secondary to primary ratios, $I_{\alpha}(E_{k})/I_{\alpha'>\alpha}(E_{k})$, such as the B/C ratio: for sufficiently high rigidities, where $\frac{X(R)}{X_{cr,\alpha}}\ll 1$, the ratio has the scaling $\propto X(R)$, hence the measurement of such ratios provides information on the grammage $X(R)$ traversed by CRs, and indirectly on the diffusion coefficient, which represents the connection between the micro-physics of particle scattering and the macro-physics of CR transport. It is worth stressing that this scaling is based on the assumption that the spallation cross sections remain energy independent even at high energies. Moreover, the assumption that the flux of a given secondary nucleus is dominated by one primary is usually bad: taking into account the contribution of all nuclei with $\alpha'>\alpha$ leads to decrease the estimate of the $X(R)$ necessary to reproduce the flux of a given secondary nucleus and therefore leads the approximate proportionality of the ratio to $X(R)$ verified at lower rigidity. Despite this, the secondary to primary ratios can be considered as a good approximation to $X(R)$ only for $R\gtrsim 100$ GV. 

The considerations above remain qualitatively valid for other secondaries such as positrons ($e^{+}$) and antiprotons ($\bar p$), with some noteworthy differences: 1) the transport of positrons and electrons may be dominated by radiative energy losses; 2) the cross section for production of antiprotons depends on energy and is known rather poorly at energies above $\sim 100$ GeV (see \cite{2014PhRvD..90h5017D} for a recent calculation).

Recently, violations of the general trend that the secondary to primary ratios decrease as $X(R)$ have been measured in the case of the positron ratio ($e^{+}/(e^{-}+e^{+})$) and the $\bar p/p$ ratio have been interpreted as problematic and potentially leading to radical changes in the standard picture illustrated above. We will discuss these implications and their limitations later in this paper.


While the recent {\it anomalies} concern relatively high energy measurements, it has long been known that the observed low energy behaviour of these ratios typically requires that either the diffusion coefficient has a plateau at rigidity $\lesssim 3 GV$, or the injection spectrum flattens at low energies and substantial re-acceleration occurs, or, finally, advection plays a prominent role (see for instance \cite{2001ApJ...547..264J}). In standard approaches to CR transport, these instances are accommodated by imposing breaks in either the diffusion coefficient or the injection spectrum. However it is worth keeping in mind that breaks in power laws are usually a signature of some relevant physical scale that enters the problem. A couple of very different scenarios that would both manifest themselves with the appearance of breaks similar to those necessary to fit the data will be discussed in \S\ref{sec:Gal-wind}.

Before concluding this section, let us recall that a residual grammage is also often introduced as a free parameter, to be added to $X(E)$. The residual grammage is usually assumed to be rigidity independent and to represent the potential contribution to secondary production inside the sources of CRs, although in some alternative scenarios of CR transport \citep{2010PhRvD..82b3009C,2014ApJ...786..124C,2016ApJ...827..119C} most grammage is assumed to be accumulated near the source rather than during propagation through the Galaxy, and an energy independent grammage at high energy is instead assumed to be the contribution of the Galaxy. Physically, the introduction of a residual grammage implies that the secondary-to-primary ratios should become roughly constant at sufficiently high rigidity. 

The simple model of CR transport illustrated above may be easily generalized to mildly more complex situations, such as a $z$-dependent non-separable diffusion coefficient (see for instance \cite{2002ApJ...572L.157D,2012ApJ...752L..13T}). 

\section{Self-excited Alfv\'en waves and their damping}
\label{sec:waves}
An aspect of CR research that has received increasingly more attention in recent years is the exploration of the non-linearities entailed by CR acceleration and propagation. The emerging picture is one in which CRs are far from passive spectators of their acceleration and transport through the Galaxy. They rather affect the environment that hosts them in two main ways: through an actual dynamical action where their energy density is large enough to compete with that of the background gas (as e.g. close to a blast wave that is providing efficient acceleration), and through their ability at generating hydromagnetic waves every time their number density is large enough (a condition that is easily realised even in the general galactic environment, far from the sources, at low enough energies). Indeed the proposal that self-generation of waves could play a fundamental role for the isotropization and confinement of CRs in the Galaxy goes back to the '70s \cite[]{skilling,holmes}, when it was first realised that super-alf\'enic streaming of CRs induces the growth of magnetic waves (Alfv\'en waves) through streaming instability. The general idea is as follows: as soon as there are energetic particles streaming at a speed that is larger than the local Alfv\'en speed (the characteristic speed of magnetic disturbances parallel to the pre-existing magnetic field in a medium), Alfv\'en waves of appropriate wavelength become unstable and grow in amplitude. This phenomenon is what is referred to as {\it streaming instability} and has long been suspected to play a major role both in CR transport across the Galaxy (see \cite{cesarsky,wentzel} for reviews) and in providing the efficient confinement that is needed in CR accelerating sources in order to reach the highest observed energies (see \cite{2013A&ARv..21...70B} and \cite{2014IJMPD..2330013A} for recent reviews).

Attention towards this instability in both contexts has been revived in recent years due to different reasons. As far as particles' confinement in the accelerators is concerned, it was realised by \cite{2004MNRAS.353..550B} that a branch of this instability, that had not previously been considered, has an especially fast growth in the vicinity of a SNR shock and can in principle explain the large magnetic fields implied by X-ray observations \cite[]{2012A&ARv..20...49V} and possibly justify CR acceleration up to the {\it knee}. Bell's initial suggestion gave rise to a wealth of studies ({\it e.g.} \citet{ziraptu08,reville08,ab09,zweibeleverett,riquelme}) on this branch of the instability (Bell's instability or non-resonant streaming instability, hereafter), aimed at further exploring its consequences in different contexts. 

At the same time the {\it classical} version of the streaming instability has also gained renewed attention in light of the breaks in CR spectra highlighted by PAMELA, AMS-02 and CREAM. The growth of this subset of unstable modes is induced by fast streaming particles at wavelengths that are resonant with their Larmor radius $r_L$, namely $k=1/r_L(p)$, where $k$ is the wavenumber of the perturbation. Resonant modes are always unstable to growth as long as the particles are super-Alfv\'enic with respect to a pre-existing magnetic field $B_0$. When this condition is satisfied, the resonant mode grows with a rate:
\be
\Gamma_{\rm CR}^{\rm RES}(k)= \frac{\pi^2}{c} \frac{v_A}{B_0} J_{\rm CR}^{\rm RES}(k)\ ,
\label{eq:resgrowth}
\ee
where $J_{\rm CR}^{\rm RES}$ is the current carried by particles (for ease of notation we here consider protons and only in the end generalise the result to include different nuclear species) with momenta $p \ge p_{\rm res}(k)=e B_0/ck$, along the direction of the background magnetic field $\vec B_0=B_0\vec z$.

The non-resonant mode has more stringent requirements for its growth, in particular the particle drift must occur at a speed 
\be
v_d>cU_B/U_{\rm CR}\ ,
\label{eq:nrcond}
\ee
where $U_{B}$ and $U_{CR}$ are the energy density in magnetic field and accelerated particles respectively.

In this case the instability has its maximum growth rate at wavenumber $k_c= 4 \pi J_{\rm CR}/(cB_0)$, at which it grows with a rate:
\be
\Gamma_{\rm CR}^{\rm NR}(k_c)= \frac{4\pi}{c} \frac{v_A}{B_0} J_{\rm CR}\ .
\label{eq:nrgrowth}
\ee
where $J_{\rm CR}$ is the total electric current carried by CRs of all momenta. 

The expressions above make immediately clear that the non-resonant instability should be important where the CR density and/or velocity are especially high, and hence within CR sources or in their vicinity, whereas for particle transport in the Galaxy, far from the sources, the resonant branch is bound to be the relevant one. 

For the purpose of the present discussion, a more useful way of writing Eq.~\ref{eq:resgrowth} is by expressing the resonant CR current in terms of particles' density gradient:
\be
J_{\rm CR}^{\rm RES}=4 \pi e \left[D(p) p^3\frac{\partial f}{\partial z}\right]_{p=p_{\rm res}}\ ,  
\label{eq:jtograd}
\ee
where $f$ is the particle density in phase space and $D(p)$ is the diffusion coefficient. The latter can be expressed through quasi-linear theory as:
\be
D(p)=\frac{4}{3\pi} \frac{r_L(p)v(p)}{kW(k)}
\label{eq:diff1}
\ee
where $kW(k)$ is the energy density of the perturbations of wavelength $k$ relative to the background magnetic field, namely, $W(k)$ is such that:
\be
\int_{1/L}^\infty W(k')dk'=\delta B^2/B_0^2\ ,
\label{eq:wknorm}
\ee
where $L$ is the largest scale in the spectrum.

Using Eq.~\ref{eq:diff1} into Eq.~\ref{eq:jtograd}, we can finally write the growth rate of the instability in a form that is useful for the back of the envelope estimates to follow:
\be
\Gamma_{\rm CR}^{\rm RES}(k)=\frac{\pi^2 e}{B_0} \frac{v_{\rm A}}{c} \left[ \frac{D(p)}{H} 4 \pi p^3 f(p)\right]_{p=p_{\rm res}}\ ,
\label{eq:gammacr1}
\ee
where we have approximated the gradient of the distribution function as $\partial f/\partial z\approx f/H$, with $H$ the size of the halo.

The magnetic perturbations will effectively grow only if the rate in Eq.~\ref{eq:gammacr1} is faster than the damping processes at work. The main damping mechanisms traditionally considered for these waves are non-linear damping (NLD hereafter) and ion-neutral damping (IND hereafter). The effectiveness of the two depends on the local ISM properties.
NLD consists in wave-wave interactions that lead the development of a turbulent cascade. A widely used prescription for the rate at which it occurs is \citep{2003A&A...403....1P}:
\be
\Gamma_{\rm NLD}(k)=(2 c_K)^{-3/2} k v_A \left(kW(k)\right)^{1/2}\ , \, \, \, \,  c_K\approx 3.6\ .
\label{eq:gnlld}
\ee
The above expression is equivalent to assuming a Kolmogorov description of the turbulent cascade. 

\begin{figure}
\includegraphics[width=\textwidth]{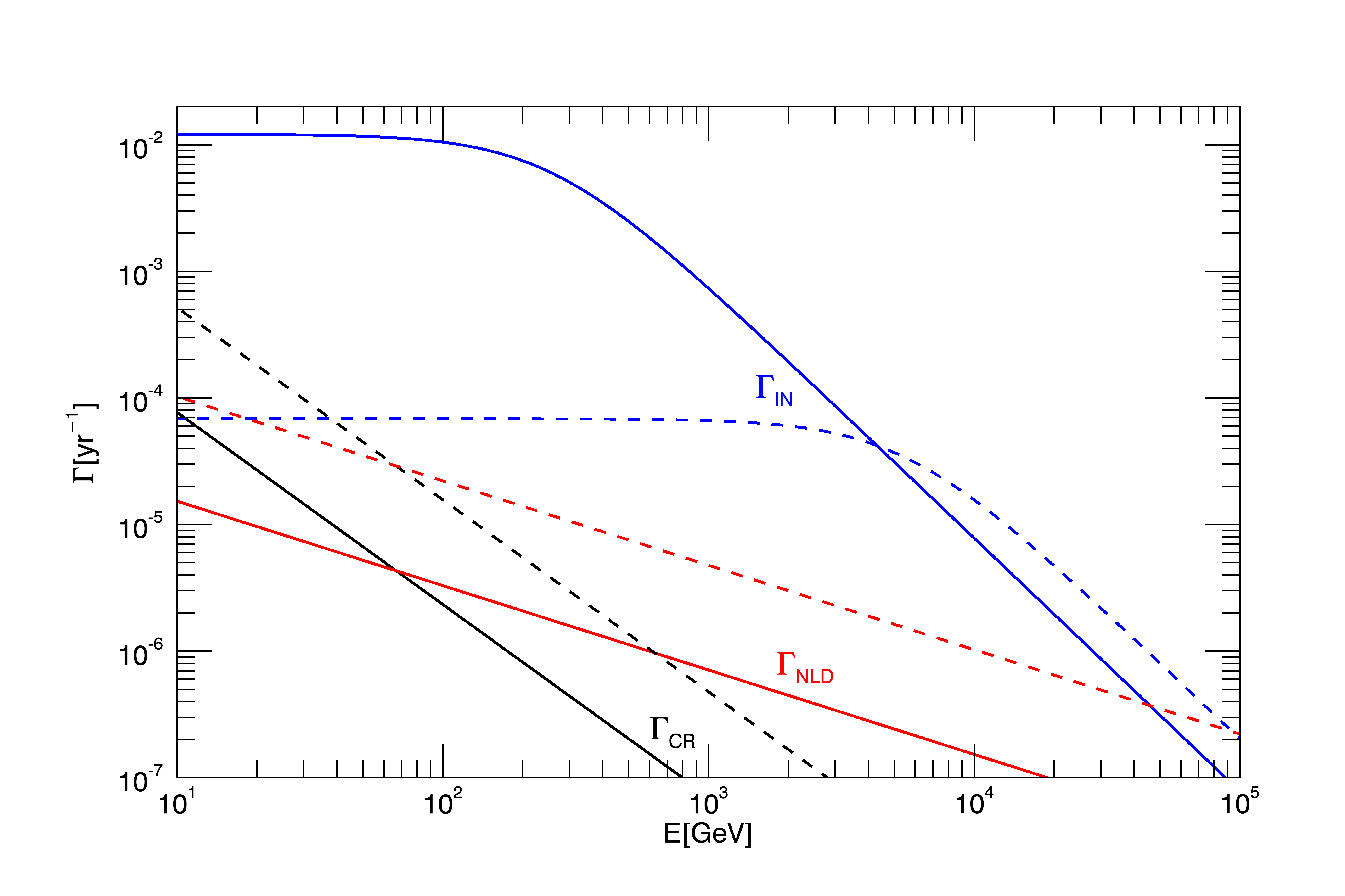}
\caption{The non-linear damping rate ($\Gamma_{\rm NLD}$, red curves), the ion-neutral damping rate ($\Gamma_{\rm IND}$, blue curves) and the growth rate of CR induced resonant streaming instability ($\Gamma_{\rm R}$, black curves) are plotted (in units of ${\rm yr}^{-1}$) as a function of particle energy (in GeV). The two sets of curves correspond to different choices of the parameters values: the solid curves are for waves in the warm, partially ionised medium ($T_4=0.8$, $B_0=3 \mu G$, $n_H=0.05 {\rm cm}^{-3}$, $n_i=0.5 {\rm cm}^{-3}$). $\Gamma_{\rm CR}$ has been computed through Eq.~\ref{eq:gammacr1} with $D(p)=1.33\times 10^{28} (H/{\rm kpc}) (pc/3 GeV)^{1/3}$ and $n_{\rm CR}$ is the proton spectrum provided by AMS02 \citep{2015PhRvL.114q1103A}. The dashed blue curve representing the IND rate corresponding to a density of neutral gas as low as $n_H=10^{-5}\ {\rm cm}^{-3}$. }
\label{fig:gammas}
\end{figure}

As far as IND is concerned, the damping rate is given by (e.g. \cite{drury96}):
\be
\Gamma_{\rm IN}(k)=\frac{k^2}{k^2+k_c^2}\nu_{\rm IN}
\label{eq:ind}
\ee
where $\nu_{\rm IN}=8.4\times10^{-9} {\rm s}^{-1} T_4^{0.4} n_H$ is the ion-neutral collision frequency, with $T_4$ the temperature of the ambient medium in units of $10^4K$ and $n_H$ the density of neutrals in units of $cm^{-3}$. The critical wavenumber separating the two regimes is $k_c=(\nu_{\rm IN}/v_A)(n_i/n_H)$, where $n_i$ is the density of ionised material. For typical ISM conditions ($T_4=1$, $B_0=3 \mu G$,  $n_i=0.45\ {\rm cm}^{-3}$ and $n_H=0.05\ {\rm cm}^{-3}$ \citep{ferriere01}) the critical wavenumber turns out to be $k_c=4.5 \times 10^{-15} {\rm cm}^{-1}$, corresponding, through the resonance condition to a particle rigidity $R_c\approx 200$ GV. At rigidities much lower than $R_c$ (wavenumbers $k>>k_c$) wave damping occurs at a rate $\nu_{\rm IN}$, while becomes progressively less important at higher energies (smaller wavenumbers) scaling as $\nu_{\rm IN}(k/k_c)^{2}$. Physically, this is due to the fact that in the presence of high frequency disturbances, ions and neutrals do not have time to become effectively coupled, so that only the ions are involved in the wave motion and in every charge-exchange event energy must be drained from the wave by the new charged particles that needs to settle down. The relative importance of IND and NLD depends on the local abundance of neutral atoms. For typical values of the parameters, as reported above, the neutral friction easily suppresses the growth of waves, as is clearly seen from the plot in Fig.~\ref{fig:gammas}, where the rates of ion-neutral damping (Eq.~\ref{eq:ind}), non-linear damping (Eq.~\ref{eq:gnlld}) and CR induced wave growth (Eq.~\ref{eq:gammacr1}) are compared. The curves in the figure refer to different conditions in the ISM (as described in the caption) and clearly show how severe a limit for wave-growth ion-neutral damping provides.

The issue of the role of IND for CR propagation in the Galaxy is an old one, dating back to the work of \cite{skilling} and \cite{holmes}, who pointed out that IND would induce a wave free zone above and below the Galactic disc (where the density of neutrals is the highest) where CRs would move ballistically. Particle isotropization would then rely on the existence of a high altitude, fully ionised region, where wave growth would be faster than IND, thereby allowing CR diffusion.
As we discuss in the following section, this idea of particle self-confinement provides a very natural way of explaining the breaks detected in the spectra of nuclei in the few hundred GeV energy range.

Another obvious place where the ion-neutral friction can be overcome is in the vicinity of the sources where $\Gamma_{\rm CR}$ is larger thanks to a larger gradient in the CR density than adopted for the plot in Fig.~\ref{fig:gammas}. We will also shortly mention this process and its consequences in \S~\ref{sec:grammage}.

\section{Galactic CR transport in self-generated waves}
\label{sec:Gal-self}
The problem of explaining what kind of turbulence would ensure effective CR scattering dates back to the very early days of CR research. On one hand, one could assume that turbulence is injected, for instance by supernova explosions and eventually cascades towards the small scales relevant for particle scattering. On the other hand, it is difficult to imagine that such turbulence would exist at high altitude above the Galactic disc. Let us consider the simplest possible approach to the problem, namely a one-zone model in which the turbulence is injected in the whole Galaxy at a rate:
\be
q_W(k)=\eta_B \delta(k-1/l_c)
\label{eq:turbinj}
\ee 
with $l_c\sim 50-100$ pc and $\eta_B$ defined such that $\eta_B=\delta B^2/B_0^2=\int dk W(k)$ (see Eq.~\ref{eq:wknorm}). This turbulence is then believed to cascade to smaller scales and a commonly (though not universally) accepted description of the cascading process is that leading to a Kolmogorov's spectrum $W(k)\propto k^{-5/3}$. In the absence of neutrals providing additional friction, the evolution of the wave spectrum under the action of cascading in $k$-space and wave excitation by the streaming CRs can be described as:
\be
\frac{\partial}{\partial k}\left[ D_{kk} \frac{\partial W}{\partial k}\right] + \Gamma_{\rm CR}W = q_{W}(k), 
\label{eq:cascade}
\ee
where $D_{kk}=k^2\Gamma_{\rm NLD}(k)$ is the diffusion coefficient of waves in $k$-space for a Kolmogorov phenomenology, with $\Gamma_{\rm NLD}$ given by Eq.~\ref{eq:gnlld}. The description of wave cascading as a process of diffusion in $k-$space was first put forward by \cite{zhou} and later applied by \cite{miller} to the description of solar flares as a second order acceleration process. Such a formalism should be considered as a mathematically convenient way to treat this process though certainly not retaining all the complexity of the phenomenon. For instance it does not include (and we do not consider here) the fact that cascading of Alfv\'enic modes proceeds in an anisotropic manner, moving power from the parallel to the perpendicular wavenumbers \cite[]{bill,gs}. In fact this represents one of the most problematic aspects of scattering theory nowadays, since modes with wavenumbers perpendicular to the regular magnetic field are in general inefficient at scattering particles. Hence, although it is a fact that scattering takes place in the Galaxy, its origin is not completely clear. It should be noticed however that, while anisotropic cascading appears to be a problem when power is mainly injected at one (typically large) scale (energy containing scale), it is expected that it is less of a problem for self-generated turbulence, which is injected continuously at all relevant scales.

A back of the envelope estimate \citep{2012PhRvL.109f1101B} of the transition scale between the dominance of CR induced turbulence and general MHD turbulence in the Galaxy can be obtained as follows. One can solve Eq.~\ref{eq:cascade} separately for the CR induced turbulence $W_{\rm CR}$ and for the externally injected turbulence $W_{\rm ext}$. The second simply evolves in a Kolmogorov's spectrum with normalisation $\eta_B$, while the first is found as the solution of $\Gamma_{\rm CR}=\Gamma_{\rm NLD}$ computed for the high energy proton spectrum measured by PAMELA and AMS-02. The spatial scale $k_{\rm tr}$ at which $W_{\rm ext}(k_{\rm tr})=W_{\rm CR}(k_{\rm tr})$, corresponds to a change in the wave spectrum that reflects into a change of the equilibrium particle spectrum at the resonant energy 
\be
E_{\rm tr}\approx 230 {\rm GeV} \left(\frac{R_{d,10} H_3^{-\frac{1}{3}}}{\xi_{0.1}E_{51} {\cal R}_{30}}\right)^{\frac{3}{2(\gamma_p-4)}}\ B_{0,\mu}^{\frac{(2\gamma_p-5)}{2(\gamma_p-4)}}\ ,
\label{eq:ear}
\ee
where $R_{d,10}$ and $H_3$ are the radius of the Galactic disk and halo in units of 10 and 3 kpc respectively, $\xi_{0.1}$ is the CR acceleration efficiency per SNR in tens of percent, $E_{51}$ is the SN explosion energy in units of $10^{51}$ erg and ${\cal R}_{30}$ is the SN rate
in units of 1/(30 yr); $B_{0,\mu}$ is the average strength of the Galactic magnetic field in $\mu$ G and finally $\gamma_p$ is the observed proton spectral index (in momentum) at energies above 300GeV (from AMS-02 data $\gamma_p=4.72$). The turbulence spectrum will be dominated by CR induced turbulence at scales smaller than $\lambda\approx r_L(E_{\rm tr})=10^{15}$ cm and by external turbulence at larger scales. As a result, the diffusion coefficient will change its energy dependence around that energy. At higher energies, particle transport is sensitive to the Kolmogorov-like part of the turbulent spectrum, $W(k) \propto k^{-\frac{5}{3}}$, and hence $D(p)\propto E^\frac{1}{3}$ (see Eq.~\ref{eq:diff1}). At lower energies the growth of waves induced by CR streaming is the dominant effect: the spectrum of turbulence acquires a stronger dependence on the wavenumber and the spectrum of particles resulting from transport becomes correspondingly steeper. The transition from a steeper spectrum at lower energies to a harder spectrum at higher energies compares extremely well with the spectral breaks measured by PAMELA and AMS-02, such transition taking place at energy $\sim E_{\rm tr}$.

A more thorough investigation of the non-linearities involved in the physics of propagation was carried out by \cite{2013JCAP...07..001A} who included in the study not only protons, but also all other nuclear species that are relevant for wave growth and, even more important, allow one to compare the theoretical predictions of the model with other observables, such as the B/C ratio. Again the waves and particles' spectra were computed self-consistently through a generalization of the same set of equations. The expressions for the wave growth and for the diffusion coefficient are generalized as follows:
\be
\Gamma_{\rm cr}(k)=\frac{16 \pi^{2}}{3} \frac{v_{\rm A}}{k\,W(k) B_{0}^{2}} \sum_{\alpha} \left[ p^{4} v(p) \frac{\partial f_{\alpha}}{\partial z}\right]_{p=Z_{\alpha} e B_{0}/kc} ,
\label{eq:gammacrall}
\ee
and
\be
D_{\alpha} (p) = \frac{1}{3} \frac{p\,c}{Z_\alpha eB_{0}} v(p) \left[ \frac{1}{k\ W(k)} \right]_{k=Z_{\alpha} e B_{0}/pc},
\label{eq:diff}
\ee
with $Z_\alpha$ being the electric charge of nuclei of species $\alpha$ and $k=Z_{\alpha} e B_{0}/p\,c$ the resonant wave number. The nonlinearity of the problem is evident here. The diffusion coefficient for each nuclear species depends on all other nuclei through the wave power $W(k)$, but the spectra are in turn determined by the relevant diffusion coefficient. In \cite[]{2013JCAP...07..001A}, the spectra of several nuclei and the ratio $B/C$ were computed self-consistently and showed to well reproduce the existing data, including the presence of breaks in the spectra of all nuclei at rigidity $\sim 100-1000$ GV. The same calculations were later applied to the AMS-02 data \cite[]{2015PhRvL.115u1101A} and Voyager \cite[]{voyager} by \cite{2015A&A...583A..95A}, where the resulting B/C ratio was also compared with the preliminary AMS-02 data available at the time.

   \begin{figure}
   \centering 
   \includegraphics[width=6.cm]{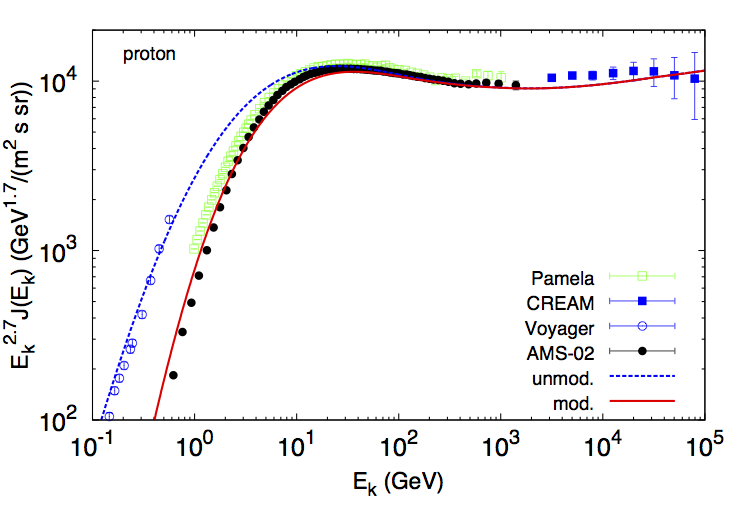}
   \includegraphics[width=6.cm]{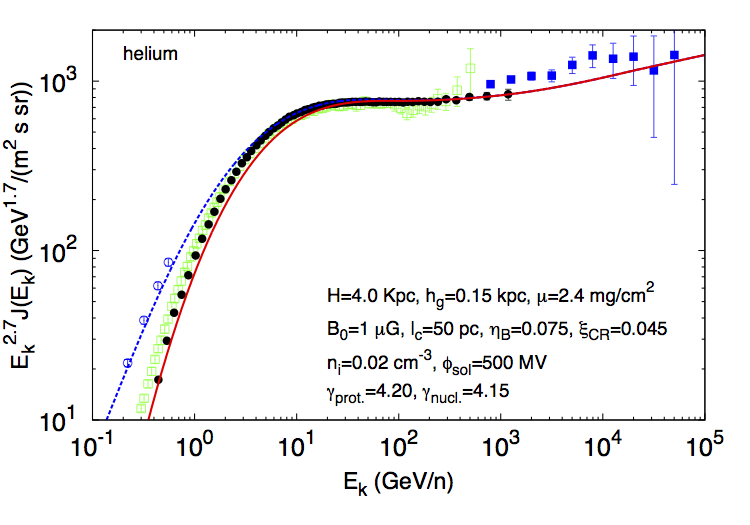}
   \caption{Spectrum of protons (left) and He nuclei (right) measured by Voyager (blue empty circles), AMS-02 (black filled circles), PAMELA (green empty squares) and CREAM (blue filled squares), compared with the prediction of the calculations (lines) by \cite{2015A&A...583A..95A}. The solid line is the flux at the Earth after the correction due to solar modulation, while the dashed line is the spectrum in the ISM.}
              \label{fig:spectrap}%
    \end{figure}

The spectra of protons and helium nuclei as calculated by \cite{2015A&A...583A..95A} are shown in Fig.~\ref{fig:spectrap}. The solid lines indicate the spectra at the Earth, namely after solar modulation modelled using the force-free approximation \cite[]{gleeson68}, while the dashed lines are the spectra in the ISM. The data points are the measurements by Voyager (empty circles, \cite{voyager}), AMS-02 (filled circles, \cite{2015PhRvL.115u1101A}), PAMELA (empty squares, \cite{2011Sci...332...69A}), and CREAM (filled squares, \cite{cream}). Fig.~\ref{fig:spectrap} shows several interesting aspects: 1) the spectra of protons and He nuclei show a pronounced change of slope at few hundred GeV/n, where self-generation of waves becomes less important than pre-existing turbulence (in fact, the change of slope takes place in rigidity). 2) The spectra calculated to optimize the fit to the AMS-02 and PAMELA data are in excellent agreement with the Voyager data (see dashed lines). This is a non-trivial conclusion, given that in this model, at sufficiently low energies (below $\sim 10$ GeV/n), particle transport is dominated by advection with self-generated waves (at the Alf\`en speed), rather than by diffusion. This fact reflects into a weaker energy dependence of the propagated spectra, which  is exactly what Voyager measured (see also \cite{potgieter2014}). It should be noticed that in order to reflect the slightly harder observed spectrum of helium nuclei at all energies, the injection spectra of hydrogen and helium used to obtain the curves in Fig.~\ref{fig:spectrap} were chosen to be $\propto E^{-2.2}$ and $\propto E^{-2.15}$ respectively. While a few models have been put forward \cite[]{2012PhRvL.108h1104M,2013ApJ...763...47P,2015ApJ...815L...1T}, a widely accepted explanation for the remarkable difference in the spectra of H and He is still missing.

\begin{figure}
   \centering 
   \includegraphics[width=8.cm]{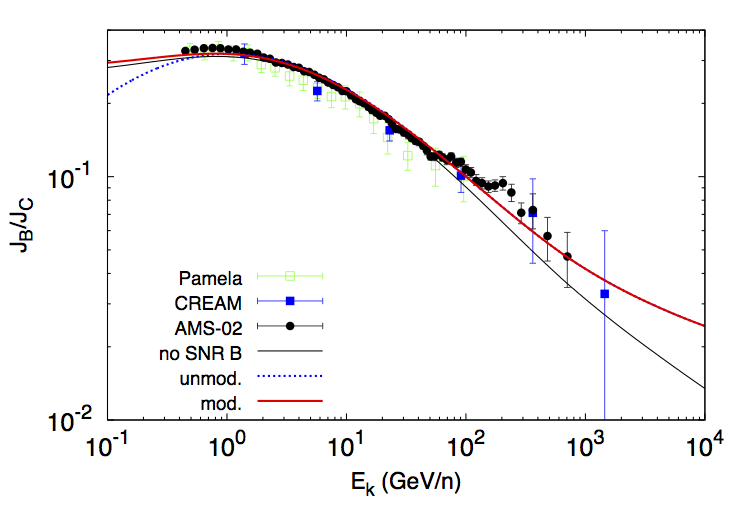}
   \caption{B/C ratio  as measured by CREAM (blue squares), PAMELA (green empty squares), and according to preliminary measurements of AMS-02 (black circles). The black/bottom solid line is the prediction of the model by \cite{2015A&A...583A..95A}, while the red/top line has been obtained by adding a source grammage of $0.15\,$g\,cm$^{-2}$, close to that given by Eq.~(\ref{eq:source}).}
              \label{fig:sec}%
    \end{figure}
    
In Fig.~\ref{fig:sec} we show the B/C ratio (solid black line) calculated by \cite{2015A&A...583A..95A} as compared with data from CREAM (blue squares), PAMELA (green squares), and the preliminary data from AMS-02 (black circles). The conclusions reached by \cite{2015A&A...583A..95A} retain their validity even in the aftermath of the newly released official AMS-02 data on the B/C ratio \cite[]{BsuC-AMS}: the model predictions fit the AMS-02 data up to $\sim 100\,$GeV/n. At higher energy, the AMS-02 analysis seems to suggest a B/C ratio somewhat higher than our prediction, which might suggest the presence of an additional contribution to the grammage traversed by CRs. The most straightforward possibility to account for such a grammage is that it may be due to the matter traversed by CRs while escaping the source, for instance, a SNR. The grammage due to confinement inside a SNR can be easily estimated as 
\be
X_{\rm SNR} \approx 1.4 r_{s} m_{p} n_{\rm ISM} c T_{\rm SNR} \approx 0.17\, {\rm g\,cm^{-2}} \frac{n_{\rm ISM}}{{\rm cm}^{-3}} \frac{T_{\rm SNR}}{2\times 10^{4}{\rm yr}},
\label{eq:source}
\ee
where $n_{\rm ISM}$ is the density of the interstellar gas upstream of a SNR shock, $r_{s}=4$ is the compression factor at the shock, and $T_{\rm SNR}$ is the duration of the SNR event (or better, the lifetime useful to confine particles up to $E\sim$TeV/n), assumed here to be of order $\sim 2\times 10^{4}$ years. The factor 1.4 in Eq.~(\ref{eq:source}) has been introduced to account for the presence of elements heavier than hydrogen in the target. Eq.~(\ref{eq:source}) is clearly only a rough estimate of the grammage at the source: first, several (in general energy dependent) factors may affect the estimate; second, it does not take into account the fact that the gas density experienced by particles changes in time during the expansion of the shell. Despite these caveats, Eq.~(\ref{eq:source}) at least provides us with a reasonable benchmark value. The solid red curve in Fig.~\ref{fig:sec} shows the result of adding the grammage accumulated by CRs inside the source to that due to propagation in the Galaxy. It is apparent that the curve accounting for production at source fits better the AMS-02 data at high rigidity, while it is also compatible with the older CREAM data.

As discussed in this section, the possibility that CRs may appreciably contribute to generating the waves responsible for their own scattering is very fascinating and pregnant with implications. Aside from the ones discussed above, one should recall a few other considerations that may be easily understood physically: first, wherever there are more sources of CRs, one should expect a higher gradient, hence more waves, which in turn result in a higher CR density. The connection between this simple line of thought and the problem of the CR gradient is intuitively clear and has been discussed recently by \cite{Recchia-grad}. Second, it can be shown in a rather general way, that self-generation of waves automatically leads to a diffusion coefficient that at rigidities below $\sim 10$ GV becomes roughly energy independent to a value $D\sim 2 v_{A}H$, where $v_{A}$ is the Alfv\'en speed and $H$ the size of the halo. The assumption that is usually made in transport calculations that the diffusion coefficient is constant in rigidity at low rigidity might in fact have a physical explanation in this non-linear effect. Even numerically, taking $v_{A}\sim 10$ km/s and $H\sim 3$ kpc, one would get $D\sim 2\times 10^{28}~\rm cm^{2}s^{-1}$, close to the value that is typically assumed in the literature to fit data.

\section{Propagation in the vicinity of sources}
\label{sec:grammage}
Near their sources, the density of CRs remains larger than the Galactic average density for quite some time, hence both the CR density and gradient are large enough that one can expect non-linear effects to dominate transport. In turn, such non-linearities act in the direction of enhancing the confinement of CRs near the sources. The picture that arises from these considerations is one in which the first stages of release of CRs into the ISM are all but trivial.


This phenomenon has been recently investigated by \cite{plesser,malkov,marta,nava1} though under somewhat different assumptions. The conclusion that is reached is however very similar: accelerated particles constitute a primary source of turbulence in the vicinity of their accelerators. The amount of turbulence that they can generate and, as a consequence, the time for which they are confined close to the sources strongly depend on the near source environment. \cite{plesser} found a semi-analytical self-similar solution of the transport equation in the assumption that NLD was the only damping process to limit the growth of the waves. A self-similar solution was also found by \cite{malkov} in the assumption that ion-neutral damping (IND) limits the growth of self-generated waves. On the other hand \cite{marta} and \cite{nava1} proposed a numerical solution of the transport equation close to a CR source in the presence of both NLD and IND. The paper of \cite{nava1} focuses on the implications of the self-generation in terms of enhanced gamma ray emission from dense clouds in the near source region.

The paper by \cite{marta} emphasizes another aspect of the problem, with potentially crucial implications for the way we derive information about particle transport from the measurement of secondary to primary ratios. Most CR sources are located in the disc of the Galaxy, where the typical density is $\sim 1~\rm cm^{-3}$ and the magnetic field is thought to be mostly ordered on scales $L_c\sim 100$ pc, possibly larger if $\delta B/B<1$ on that scale. As we already discussed in \S\ref{sec:waves}, the ionisation of the ambient medium is of crucial importance. 

In order to account, to some extent, for the uncertainty in the role of IND around sources of CRs, \cite{marta} considered the following cases: (1) No neutrals and gas density $0.45~\rm cm^{-3}$ (this is the best case scenario in terms of importance of the near source grammage);
(2) Neutral density $n_{n}=0.05~\rm cm^{-3}$ and ion density $n_{i}=0.45~\rm cm^{-3}$; (3) Neutral density $n_{n}=0.03~\rm cm^{-3}$ and ion density $n_{i}=0.45~\rm cm^{-3}$;
(4) rarefied totally ionized medium with density $n_{i}=0.01~\rm cm^{-3}$. The reasons why these particular values were chosen will become clear in the following. The calculation also assumes a SN explosion energy of $10^{51}$ erg, a coherence length of the background magnetic field $L_c=100$ pc and CR acceleration efficiency of 20 \%.

\begin{figure}
\includegraphics[width=0.8\textwidth]{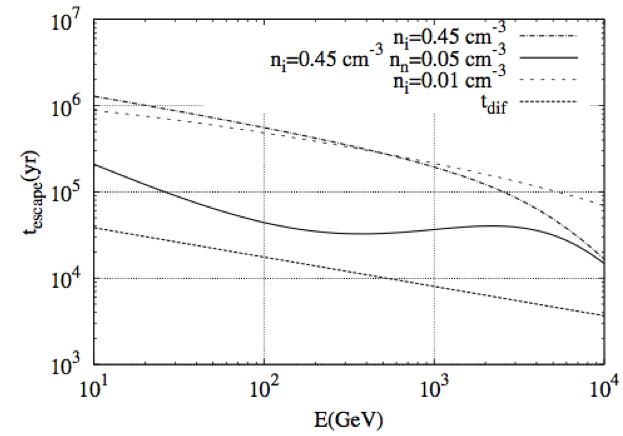}
\caption{Escape time of CRs from the near-source region. A coherence length of the background magnetic field $L_c=100$ pc is assumed. Three cases are considered with varying densities of neutrals and ions: (1) $n_n=0$, $n_i=0.45 {\rm cm}^{-3}$; (2) $n_i=0.45 {\rm cm}^{-3}$ and $n_n=0.05 {\rm cm}^{-3}$; (3) $n_n=0$, $n_i=0.01 {\rm cm}^{-3}$. The dotted line refers to the escape time calculated assuming the Galactic diffusion coefficient rather than the self-generated one. The figure is from the work by \cite{marta}.}
\label{fig:tescsrc}
\end{figure}
 
As the plot in Fig.~\ref{fig:tescsrc} (from \cite{marta}) shows, in the absence of neutrals, wave generation is so efficient that even CRs in the TeV energy range can be confined within $\sim 100$ pc from their sources for several tens of thousands of years, while in the absence of self-generation they would leave the same region in less than $10^4$ years (curve labelled as $t_{\rm dif}$ in the plot). Conversely the residence time close to the source becomes much shorter, and eventually comparable to that expected in the absence of self-generation, as the fraction of neutral hydrogen increases.

In all cases considered by \cite{marta} the residence time in the near source region remains shorter than the residence time in the Galaxy at all energies, while  the grammage accumulated during this time can be non-negligible with respect to the overall grammage, as inferred, for example, from B/C measurements. This is due to the fact that while CRs spend most of the time propagating through the tenuous gas in the galactic halo, they mostly accumulate grammage when crossing the galactic disk. If $\tau_{s}\simeq L_{c}^{2}/D_{sg}$ is the time spent within $L_c$ of the source under the action of the self-generated diffusion coefficient $D_{sg}$, then the accumulated grammage becomes comparable with the Galactic grammage as soon as $\tau_{s}\sim \tau_{H} h_{d}/H\approx 10^{-2}$, with $h_{d}$ the height of the disk. The work by \cite{marta} shows that in the absence of neutral hydrogen, responsible for IND, the self-generation of waves may be sufficient to warrant these conditions. On the other hand, the fraction of neutral hydrogen in the ISM depends on the phase of the gas: the densest gas (the one that would provide more grammage) is also the coldest and thereby the least ionised.

\begin{figure}
\includegraphics[width=0.8\textwidth]{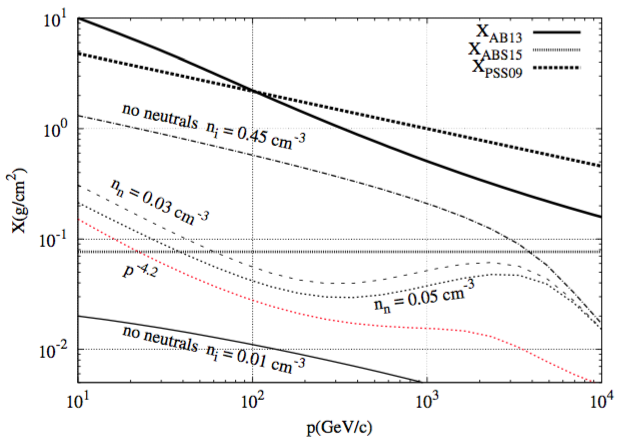}
\caption{Grammage accumulated by CRs in the near-source region for $L_c=100$ pc in the 3 cases: 1) $n_n=0$, $n_i=0.45 {\rm cm}^{-3}$; (2) $n_i=0.45 {\rm cm}^{-3}$ and $n_n=0.05 {\rm cm}^{-3}$; (3) $n_n=0$, $n_i=0.01 {\rm cm}^{-3}$ as labelled. The thin dotted (red) line corresponds to case (2) but with slope of the injection spectrum 4.2. The thick dashed line (labelled as XPSS09) shows the grammage inferred from the measured B/C ratio \cite[]{fit}, while the thick solid line (labelled as XAB13) shows the results of the non-linear propagation of \cite{2013JCAP...07..001A}. The horizontal (thick dotted) line (labelled as XABS15) is the source grammage, as estimated by \cite{2015A&A...583A..95A}. The Figure is from the work by \cite{marta}.}
\label{fig:grammage}
\end{figure}

The grammage in the near source region as calculated by \cite{marta} is shown in Fig.~\ref{fig:grammage}: in the absence of neutral hydrogen and with a gas density $n_{\rm WIM}\approx 0.45 {\rm cm}^{-3}$ (typical of the Warm Ionised Medium or WIM) the grammage accumulated by particles in the near source region is comparable with the overall value inferred from B/C measurements. However, an ionisation fraction of 100 \% typically requires a much higher temperature ($T\approx 10^6$ K as compared to $T\approx 10^4$ K) and a much lower density ($n\approx 0.01 {\rm cm}^{-3}$) than those of the WIM. In this hot tenuous phase of the ISM, as Fig.~\ref{fig:tescsrc} and \ref{fig:grammage} show, the CR confinement time close to the source is large but the accumulated grammage is negligible. When considering, instead, a source immersed in the WIM with a more realistic value of the ionisation fraction (90 \% rather than 100 \%), one finds that IND is important but does not completely eliminate the effect of self-confinement. The wave growth is diminished, but not entirely quenched as would be the case for propagation in the Galaxy. The accumulated grammage remains non-negligible.

Moreover, as argued by \cite{ferriere}, the WIM, having temperature $\sim 8000$ K, is expected to be made of fully ionized hydrogen, while helium would be partially ionized. This latter picture would have prominent consequences in terms of IND, in that this process is due to charge exchange between ions and the partially ionized (or neutral) component, but the cross section for charge exchange between H and He is about three orders of magnitude smaller \cite[]{aladdin} than for neutral and ionized H, so that the corresponding damping rate would be greatly diminished. Unfortunately, at present, there is no quantitative assessment of this phenomenon and we can only rely on a comparison between cross sections of charge exchange. On the other hand, it is also possible that a small fraction of neutral hydrogen is still present, in addition to neutral helium: following \cite{ferriere}, the density of neutral H is $\lesssim 6\times 10^{-2} n_{i}$, and for $n_{i}=0.45  \rm cm^{-3}$ this implies an upper limit to the neutral H density of $\sim 0.03~\rm cm^{-3}$ (justifying the choice of parameters in case (3) of Figs.~\ref{fig:tescsrc} and \ref{fig:grammage}).

Fig.~\ref{fig:grammage} illustrates in a clear way how the phenomenon of self-confinement of CRs close to their sources can have potentially important implications on the interpretation of secondary-to-primary ratios in terms of grammage traversed in the Galaxy. The importance of these considerations depend in a critical way on the density of gas and the density of neutral hydrogen in the regions around sources, and on the size of the regions where the Galactic magnetic field can be considered as roughly ordered. 

Given the uncertainties discussed above, it is clear that further investigation of alternative signatures of the confinement of CRs in the near source region has to be carried out, for instance aimed at determining the level of diffuse gamma ray emission from such regions. The confinement of CRs near their sources might also be of the highest importance to establish the feasibility of alternative scenarios of CR propagation, such as those discussed by \cite{2010PhRvD..82b3009C,2014ApJ...786..124C,2016ApJ...827..119C}.

Before concluding this section, however, we would like to mention another instance in which it seems unavoidable that CR self-induced turbulence plays a role, namely the case of UHECR propagation far from the sources that have accelerated them. In that case the conditions are likely such that the non-resonant, current driven mode (Eq.~\ref{eq:nrgrowth}) of the streaming instability can grow. CRs leaving their sources find themselves in the extremely low IGM magnetic field, so that given the values of luminosity that are usually quoted for the sources of UHCRs ($L_{\rm CR}\approx 10^{44}$ erg/s), the CR energy density easily exceeds the magnetic energy density according to the criterion in Eq.~\ref{eq:nrcond}. As a consequence, the non-resonant streaming instability can grow and give rise to a turbulent magnetic field that, depending on what sets the saturation, can easily by in the $nG$ range. For typical parameters the effect is sufficient to ensure confinement of low enough energy particles for times longer than the age of the Universe \cite[]{martaUHECR}. An interesting aspect of this model is that it naturally predicts a low energy cut-off in the UHECR spectrum at an energy of about $10^7-10^8$ GeV. Such a low energy cut-off is exactly what some phenomenological models of UHECR origin and propagation require.

\section{CR induced Galactic winds}
\label{sec:Gal-wind}

Galactic winds may affect star formation, through the regulation of the amount of gas available \citep{2007MNRAS.377...41C,2013MNRAS.428..129S} and inject hot gas in the galactic halo. In our Galaxy there are indications of the existence of such hot gas from X-ray observations \cite[]{1994Natur.371..774B,1999A&A...347..650B}. Most important, a Galactic wind may affect the propagation of CRs \cite{Ptuskin:1997A&A...321..434P,Recchia-wind}, and in turn CRs may contribute to launching such winds \citep{1975ApJ...196..107I,1991A&A...245...79B,2008ApJ...674..258E}. The very fact that CRs have to leave the Galaxy in order to reach some stationary condition implies that a gradient in CR density (and pressure) must exist between the disc and the escape surface. The $-\nabla P_{CR}$ force acts in the direction pointing away from the Galactic disc, while the gravitational force exerted by dark matter, baryons and stars in the Galaxy, pulls material downward. If the CR induced force prevails on the gravitational pull close to the disc, then a Galactic wind may be launched. This is however only a necessary (but not sufficient) condition for launching a wind, in that material can be lifted up and fall down in what is known as Galactic fountains. A wind is launched only if the gas can be accelerated to supersonic speeds and its properties connected smoothly with the boundary conditions at infinity. 

The dynamics of CR induced winds, where CR propagation occurs because of their interaction with self-generated waves, is extremely complex and in the past some useful assumptions were made to simplify the problem. For instance, in the pioneering work of \cite{1991A&A...245...79B}, the authors assumed that the diffusivity of CRs is vanishingly small, so as to transform the transport equation of CRs in a fluid equation. Within this assumption, the so-called wind solutions were found but no information on the spectrum of CRs was retained. Later, \cite{Ptuskin:1997A&A...321..434P} proposed a simplified way to keep information on the CR spectrum and CR spatial distribution, and included Galactic rotation. Recently, \cite{Recchia-wind} proposed an iterative semi-analytical method to determine the wind structure, the CR spectrum and spatial distribution and to determine whether a CR induced wind may be launched in the first place, depending on the conditions at the base of the wind (Galactic rotation was not included in such calculation). 

A physical understanding of the way CR induced winds work was proposed by \cite{Ptuskin:1997A&A...321..434P}, and although their conclusions are usually not confirmed quantitatively in a generic situation (see \cite[]{Recchia-wind} for a discussion), the physical intuition turns out to be correct and very useful.

The basic assumption of \cite{Ptuskin:1997A&A...321..434P} is that near the disc the advection velocity (dominated by the Alfv\'en speed) scales approximately linearly with $z$, $v_{A}\sim \eta z$. Now, it is easy to imagine that while the advection velocity increases with $z$, it reaches a critical distance, $s_{*}$, for which advection dominates upon diffusion. This happens when
\begin{equation}
\frac{s_{*}^{2}}{D(p)}\approx \frac{s_{*}}{v_{A}(s_{*})} \, \Rightarrow \, s_{*}(p) \propto D(p)^{1/2},
\end{equation}
where we used the assumption of linear relation $v_{A}\sim \eta z$ and that the diffusion coefficient has small spatial variation, which turns out to be true in self-generated scenarios. Now, when diffusion dominates, namely when $z\lesssim s_{*}(p)$, one can neglect the advection terms and make the approximate statement (as in the standard diffusion model), that $D(p)\frac{\partial f}{\partial z}|_{z=0}\approx -Q_{0}(p)/2 \propto p^{-\gamma}$. The quantity $s_{*}(p)$, that depends naturally on particle momentum, plays the role of the size of the diffusion volume and one can show that, similar to a leaky box-like model, the equilibrium spectrum in the disc is 
\begin{equation}
f(p)\sim \frac{Q_{0}(p)}{s_{*}(p)}\frac{s_{*}^{2}}{D(p)} \sim Q_{0}(p)\, D(p)^{-1/2}.
\end{equation}
In other words, $s_{*}(p)$ plays the role of $H$ in the standard model of CR transport. There are a few very important implications of this line of thought: 1) the energy dependence of the spectrum of CRs in the Galactic disc is not simply proportional to $Q_{0}(p)/D(p)$, as is usually assumed, but rather to $Q_{0}(p)\, D(p)^{-1/2}$, where however $D(p)$ is in turn a function of the CR spectrum. Using the expressions for the growth of Alfv\'en waves, Eq. \ref{eq:resgrowth}, and their rate of damping, one can demonstrate that for $Q(p)\propto p^{-\gamma}$, $D(p)\propto p^{2\gamma-7}$, so that injection spectrum and observed spectrum are directly connected, at least for the energies where diffusion dominates. These expressions hold as long as NLD is the main channel of wave damping. 2) The size H of the halo, where free escape should occur, loses its meaning: this is a positive outcome of these models since the boundary condition typically imposed at $|z|=H$ is somewhat artificial, and shows how weak the predictions of the so-called standard model of CR transport may be. The role of $H$ is played by the physical quantity $s_{*}(p)$ which however is an output of the problem (not imposed by hand) and depends on the particles' momentum. 

\begin{figure}
	\includegraphics[width=\columnwidth]{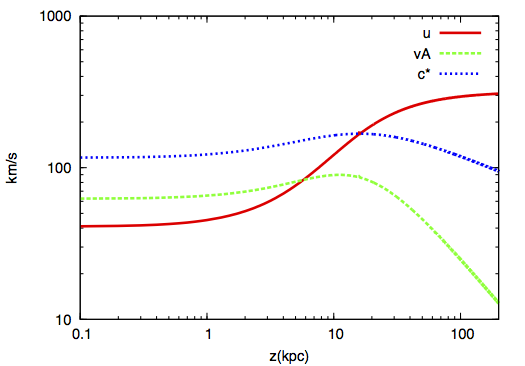}
    \caption{Wind velocity (red solid line), Alfv\'en velocity (green dashed line) and sound speed (blue dotted line) as a function of the height $z$ above the disc, for a reference case.}
    \label{fig:velocity}
\end{figure}

As pointed out above, the simple recipe proposed by \cite{Ptuskin:1997A&A...321..434P} does not apply to realistic situations because it is based on the assumption that at the base of the wind the advection velocity vanishes. In fact, the Alfv\'en speed and the wind velocity have a finite value at the base of the wind, which implies that at low energies the transport of CRs is dominated by advection. This can be seen in Fig.~ \ref{fig:velocity}, where we show the wind velocity (red solid line), Alfv\'en velocity (green dashed line) and sound speed (blue dotted line) as a function of the height $z$ above the disc for a typical wind case. 

\begin{figure}
	\includegraphics[width=\columnwidth]{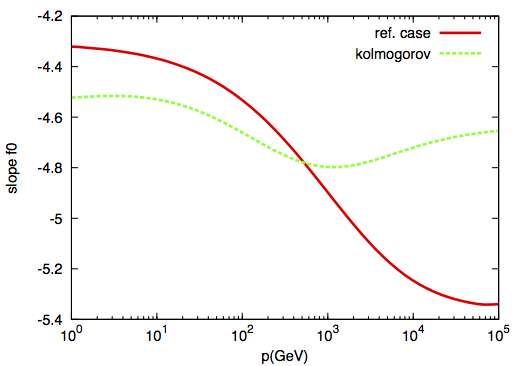}
    \caption{Slope of the CR spectrum in the disc of the Galaxy at the position of the Sun for the same case as in Fig. \ref{fig:velocity} (red solid line) and for the case in which the wind is launched at $z_{0}=1$ kpc from the disc and there is a near-disc region where the diffusion coefficient is assigned and advection is absent (green dashed line).}
    \label{fig:f0}
\end{figure}

This has important implications for the spectrum of CRs as observed at a location compatible with that of the Sun: as shown by \cite{Recchia-wind}, most wind solutions that may be found do not lead to CR spectra that are similar to the observed one, although they do correspond to dynamically feasible solutions with a substantial mass loss (of order $\sim 0.5-1~M_{\odot}~yr^{-1}$ integrated on the Galactic disc). For instance the slope of the spectrum of CRs at the Sun as calculated for the same wind model corresponding to Fig. \ref{fig:velocity} is shown as a (red) solid line in Fig. \ref{fig:f0}, clearly at odds with observations (even qualitatively). 

A rather generic problem of all transport models (including wind models) is that they are mostly based upon resonant interactions of charged particles with Alfv\'en waves. However, Alfv\'en waves are expected to be severely damped due to ion-neutral damping. This damping is so severe that within $\sim 1$ kpc from the Galactic disc, where most neutral hydrogen is located, there can hardly be any Alfv\'en waves. For this reason, in original approaches to the wind problem (e.g. \cite{1991A&A...245...79B}) the effective CR plasma coupling was guaranteed by assuming that the wind be launched at $z_{0}\simeq 1$ kpc. This near-disc region is also the one that is most perturbed by the direct action of repeated supernova explosions: for this reason, one could speculate that this region may host a local, non self-generated, turbulence directly associated to hydrodynamical turbulence. \cite{Recchia-wind} investigated the implications of this assumption in the case that the near-disc region is characterized by a Kolmogorov-like diffusion coefficient. In this scenario the base of the wind was assumed to be at $z_{0}=1$ kpc. In passing, we notice that the importance of this near-disc region was already recognized by \cite{1993A&A...269...54B}. 

The assumption of pre-existing (namely not self-generated) turbulence in the near disc region allows one to have spectra that qualitatively resemble the spectrum of CRs observed at the Earth, as shown by \cite{Recchia-wind}. This means that the low energy spectrum (below $\sim 10$ GeV) is dominated by advection with waves and is close to the injection spectrum; at energies above $10$ GeV the spectrum is fully determined by self-generated waves and has a slope close to the observed one; at energies above $\sim$ TeV, the spectrum hardens because this part of the spectrum is dominated by diffusion in the near disc region. This spectral shape, shown as a (green) dashed line in Fig. \ref{fig:f0}, is qualitatively similar to the one recently measured for protons and helium nuclei by PAMELA and AMS-02.  

It is important to keep in mind that the launching of a CR induced wind actually results from the interplay between CRs and gravity, so that the characteristics of the wind and even the possibility to launch a wind (rather than a time dependent outflow that may fall back onto the Galactic disc) depend on the gravitational potential, especially the one contributed by dark matter. A careful study of the dependence of the wind characteristics on the gravitational potential and the implications of CR induced winds in terms on mass loss from the Galaxy is required in order to have a complete picture of the applicability of wind models to the Galactic CR phenomenology.

\section{Secondary particles and antiparticles}
\label{sec:secondary}
In recent years the PAMELA and AMS-02 experiments have provided us with plenty of new results concerning the flux of secondary CR particles and antiparticles, and some of these findings are rather surprising and intriguing. The first surprising finding was the unequivocal detection by PAMELA \citep{2009Natur.458..607A} of a growing positron fraction ($e^+/(e^-+e^+)$), following earlier, but statistically less significant claims by CAPRICE \citep{CAPRICE}, HEAT \citep{HEAT} and AMS-01\citep{AMS01}. This discovery was then confirmed with even higher statistical significance by AMS-02 \citep{2013PhRvL.110n1102A}. Both PAMELA \citep{PAMELAElPos} and AMS-02 \citep{2014PhRvL.113n1102A} also showed that the increasing fraction is due to a flattening in the positron spectrum rather than a steepening of the electron distribution. 

More recently, AMS-02 has also published the results of their measurement of the flux of antiprotons and of the $\bar p/p$ ratio \citep{AMS02pbar}: at energies above $\sim 60$ GeV or so, the spectrum of antiprotons appears to be rather hard, and in fact it has been pointed out \cite[]{AMS02pbar,lipari} that it has about the same slope as the spectrum of positrons and protons. Both the rising positron fraction and the hard spectrum of $\bar p$'s appear to be in conflict with the standard picture of CR transport in the Galaxy, hence their discovery has stimulated enormous debate in the community. Below we briefly discuss these anomalies and the ideas put forward to explain them.

In the standard picture of CR transport in the Galaxy, both positrons and antiprotons are secondary products of inelastic hadronic collisions. Both the ratios $\bar p/p$ and $e^{+}/(e^{-}+e^{+})$ are then expected to decrease with energy as a result of the fact that the grammage traversed by CRs is decreasing with energy. For antiprotons, the effects of propagation are easier to describe than for leptons, in that they are not expected to suffer significant energy losses during this process. To first approximation, the ratio $\bar p/p$ should therefore decrease in the same way that the B/C ratio does, provided that the production cross section be energy-independent and that both the spectrum of the parent protons and the diffusion coefficient be pure power-laws over the entire range covering both the energies of secondaries and those (a factor of ten higher) of primaries. All these conditions are potentially violated: 1) the cross section for antiproton production does depend on energy, as recently discussed by \cite{2014PhRvD..90h5017D}; 2) the spectrum of both protons and He nuclei is characterized by spectral breaks, as discussed in \S\ref{sec:Gal-self}, with harder spectra at higher energies - moreover we stress that some contribution to $\bar p$ production is associated to interactions of He nuclei, whose spectrum is harder than that of protons; 3) as discussed by \cite{2012PhRvL.109f1101B,2013JCAP...07..001A,2015A&A...583A..95A,2012ApJ...752L..13T}, the reason for harder nuclear spectra may reside in a change of slope of the diffusion coefficient experienced by particles, hence also the last of the above mentioned conditions might be violated. 

In addition, the recently published AMS-02 results concerning the B/C ratio \cite[]{BsuC-AMS} suggest that the dependence on energy of the grammage is relatively mild, $X_{\rm esc}(E)\propto E^{-1/3}$, in agreement with a Kolmogorov's description of turbulence in the Galaxy (although see discussion in \S \ref{sec:Gal-self}). The combination of all these effects leads us to expect that perhaps the anomaly on the $\bar p/p$ ratio might in fact be less severe than initially thought. In fact, even some current studies \cite[]{2015JCAP...12..039E,2015JCAP...09..023G} reached the same conclusion by just taking into account the uncertainties in the cross section for $\bar p$ production and those associated to CR transport.

As far as leptons are concerned, the equilibrium spectrum is made slightly more complicated by the (presumably) non-negligible losses. If we take the standard value of magnetic and radiation energy density in the Galaxy, $U_B\approx U_{\rm rad}\approx 1$ eV cm$^{-3}$, the radiative loss time, $\tau_{\rm loss}(E)\approx 2.5\times 10^8$ yr $(E/{\rm GeV})^{-1}$, is shorter than the propagation time $\tau_{\rm esc}(E)\approx 2 \times 10^8$ yr $(E/{\rm GeV})^\delta$ already at $E>$ few GeV, for $0.2<\delta<0.8$ as inferred from recent studies of CR propagation (e.g. \cite{trotta}). In this case, in the energy range of our interest, the equilibrium spectrum of electrons, which we approximate as all primaries, is described by:
\be
N_{\rm e^-}(E)\propto E^{-\gamma_{\rm inj,el}}\ E^{-\beta}\ ,
\label{eq:elspec}
\ee
where the first term on the {\it rhs} comes from injection and the second from propagation. The form of the latter is determined by the combination of diffusive propagation and energy losses that limit the number of contributing sources with respect to the case of nuclei, thereby imposing constraints in terms of distance in space and time: in the case of a spatially independent diffusion coefficient that scales as $E^\delta$ and radiative (synchrotron and Inverse Compton) losses with rates that scale as $E^{2}$, one finds $\beta=(1+\delta)/2$.

Analogously, for positrons, if they are all secondaries, we find:
\be
N_{\rm e^+}(E)\propto E^{-(\gamma_{\rm inj,p}+\delta)}\ E^{-\beta}\ ,
\label{eq:posspec}
\ee
where what appears in the first term of the {\it rhs} is the equilibrium proton spectrum ($\propto E^{-\gamma_p}$).
The result for the positron fraction is then:
\be
\chi=\frac{\Phi_{e^+}}{\Phi_{e^+}+\Phi_{e^-}}\approx\frac{\Phi_{e^+}}{\Phi_{e^-}}\propto E^{-(\gamma_{\rm inj,p}-\gamma_{\rm inj,el}+\delta)}\ ,
\label{eq:chiexp}
\ee
and in the standard scenario, with $\gamma_{\rm inj,p}=\gamma_{\rm inj,el}$, $\chi\propto E^{-\delta}$ is expected to monotonically decrease with energy. One can easily check that the decreasing trend is confirmed even in situations in which the diffusion coefficient does not have a perfect power law dependence on energy. The generality of the result explains why the observed increase was found to be puzzling and stimulated a plethora of attempts at explaining it.  These attempts proceeded in three different directions: 1) dark matter inspired models; 2) modifications of the standard model to accommodate the observed anomalies; 3) astrophysical sources of the excess positrons. A comprehensive review of these models was presented by \cite{2012APh....39....2S}.

In the following we will summarize the models that invoke modifications of CR transport and those based on astrophysical sources of the positron excess.

\subsection{Non-standard propagation models}
\label{sec:nonstandard}

A few models were recently introduced that questioned the pillars of the standard model of CR transport, suggesting that the observed positron excess is in fact the result of the failure of one or more of the assumptions that are usually made in treating the transport of CRs throughout the Galaxy. Two such attempts were published in the last few years by  \cite{katz10,katz13} and by \cite{lipari}. The former work, that preceded the measurement of the B/C ratio by AMS-02, started from considering the B/C ratio as a good indicator of the grammage, $X_{\rm esc}\propto E^{-1/2}$, and illustrates the calculations of the positron (and antiproton) flux as due to inelastic $pp$ collisions, corresponding to such grammage. The authors find that the flux of the observed high energy positrons is saturated by their predicted flux, hence they conclude that the hard spectral shape of positrons is to be attributed to energy losses, more important at low energies than at the highest in their approach. This rather unusual finding is justified by the authors by assuming that synchrotron losses are rather low (interstellar magnetic field around 1 $\mu$G) and the radiation field, with energy density around 1 eV/${\rm cm}^3$, contains a dominant UV component, so that losses are reduced for high energy particles due to Klein-Nishina effects. In this approach it is therefore suggested that the whole flux of positrons is due to standard CR interactions and that the alleged excess is actually due to a poor understanding of CR transport. In order to reproduce the data the authors also suggest that one should consider an energy dependent halo size. 


On the other hand, the work of \cite{lipari} adopts antiprotons and positrons as the starting point. The author notices that the $e^{+}$ and $\bar p$ production rates are in perfect accord with those calculated from standard CR interactions in the ISM and that their spectral shape is consistent with being the same as that of protons at energies above $\sim 300$ GeV, thereby strengthening the conclusion that the observed flux of antiprotons and positrons is completely explained in terms of CR interactions. 

In the standard model of CR transport, where the grammage is inferred from the B/C ratio, the propagation of leptons (electrons and positrons) is dominated by radiative losses for energies above $\sim 10$ GeV. The effect of losses is such as to steepen the spectrum of leptons and lead to a spectral difference with antiprotons. It follows that the conclusion of \cite{lipari} can only be considered as potentially viable if energy losses are negligible, which requires a residence time in the Galaxy much smaller than usually assumed. This also implies that the typical grammage traversed by the CRs responsible for production of $e^{+}$ and $\bar p$ must be smaller than that inferred from measurements of the B/C ratio: in other words, in this scenario the boron production must be decoupled from the production of other secondaries.


One scenario that would fit these requirements is the so called ``nested leaky box'' (NLB) scenario suggested by \cite{2010PhRvD..82b3009C,2014ApJ...786..124C,2016ApJ...827..119C}. The basic assumption of the NLB model is that CRs accumulate most grammage in {\it cocoons} around the sources, while lesser grammage is accumulated during propagation throughout the Galaxy. The former grammage is assumed to be energy dependent while the latter is assumed to be energy independent. The two values of the grammage become comparable around few hundred GeV/n. Since secondary Boron nuclei are produced by primary CRs (mainly Carbon and Oxygen) at the same energy per nucleon, the B/C ratio reflects the energy dependence of the grammage at the same energy, hence one should expect that the B/C should be decreasing with energy (reflecting the near source grammage) at $\lesssim 100$ GeV/n and become energy independent at higher energies. Since the $e^{+}$ and $\bar p$ production is characterized by a large inelasticity ($E_{e^{+},\bar p}\sim 0.1 E_{p}$ with $E_{p}$ the proton energy), the production of $e^{+}$ and $\bar p$ at energy $\sim 10-100$ GeV reflects the grammage  traversed by CR at $0.1-1$ TeV, where the grammage is assumed to be flat in the context of the NLB model. One should keep in mind that the recent paper by AMS-02 on the measurement of the B/C ratio up to $\sim 1$ TeV/n did not find evidence for a flattening of the ratio at the highest energies. The idea of cocoons around sources, presented by \cite{2010PhRvD..82b3009C} as a speculative possibility, might be realized in Nature due to the mechanism of CR self-confinement discussed in \S~\ref{sec:grammage} (see work of \cite{marta}), although, as discussed above, this scenario also encounters several difficulties.


\subsection{Additional astrophysical sources}
\label{sec:pulsars}
Here we assume that CR propagation through the Galaxy occurs in the most standard and simplest way, namely with leptons loosing energy radiatively and all particles propagating with an energy dependent diffusion coefficient, approximately uniform in space. In this scenario there is no doubt that the positron excess requires additional sources. 
A number of sources of positrons or physical processes that lead to excess production of positrons have been put forward. Among these, two have attracted much attention: 1) the model of \cite{2009PhRvL.103e1104B} that suggested that positrons (and electrons) produced as secondaries inside the acceleration region of primary CRs also participate in the acceleration process and acquire a hard spectrum. 2) Pulsars, which are ``natural'' sources of positrons and can explain the observed excess \cite[]{2009JCAP...01..025H,2012CEJPh..10....1P}: we used the wording``natural" because the nature of pulsars as positron factories has been well established for a very long time, soon after their discovery and since the first attempts at describing their magnetospheres. In fact even their contribution to the CR positron flux had been previously studied by \cite{1995A&A...294L..41A}.

The model of \cite{2009PhRvL.103e1104B} also implies that an excess of antiprotons \cite[]{2009PhRvL.103h1103B} and boron \cite[]{2009PhRvL.103h1104M} should appear at energies $\gtrsim 100$ GeV/n. Although the situation of antiprotons, as discussed above, is not clear, we can say with confidence that no rise in the B/C ratio has been observed by AMS-02, which should be considered as a strong argument against such model.

As for pulsars, in the vicinity of such an object the electromagnetic fields are so intense that pair cascades develop with very high multiplicity: each electron extracted from the star surface typically produces $\sim 10^{4}-10^{6}$ $e^+$-$e^-$ pairs. Aside from being theoretically predicted, direct evidence of this phenomenon is provided by multi-wavelength observations of Pulsar Wind Nebulae, bright synchrotron and IC nebulae surrounding many young pulsars (see {\it e.g.} \cite{amato14} for a review). Since these particles will have to be released in the ISM at some point, their contribution to CR leptons is unavoidable and must be taken into account in any model aimed at explaining the positron excess. In the following we try to provide a realistic estimate of its importance.

The $e^+$-$e^-$ pairs created in the pulsar magnetosphere become part of the relativistic wind into which pulsars convert most of their rotational energy. The interaction between the wind and the surrounding medium, the SNR during early stages and the ISM later on, is what makes the PWN shine: a shock develops from this interaction and propagates towards the pulsar down to a distance that guarantees pressure equilibrium between the unshocked wind and the downstream nebula. Extremely efficient particle acceleration occurs at this shock: long power-law spectra extending from about 1 GeV to even 1 PeV are formed and the radiation of these particles in the ambient magnetic field directly reveals the accelerated particle spectrum. The $e^+$-$e^-$ pairs are seen to be described by a flat spectrum ($N(E)\propto E^{-\gamma}$ with $1<\gamma<2$), at low energies, which then steepens to $\gamma>2$ beyond a few hundred GeV. The hard lepton spectrum at low energies is extremely appealing in terms of explaining the CR positron spectrum if, as expected, at some point in the history of the PWN these particles are released into the ISM. What needs to be assessed is the effective rate of release of the pairs and their potential contribution to the CR spectrum. 

At the energies that are relevant to explain the excess, the majority of the pairs cannot leave the PWN for as long as the pulsar is embedded in the parent SNR. However, pulsars are typically characterized by high proper motion (with an average speed of 500 km/s) and they are expected to break out of the remnant in a few $\times 10^5$ yr for typical parameters. At this point, their supersonic motion through the ISM transforms them in Bow Shock PWNe. The plasma in front of the pulsar is now confined by the bow shock, but particles flowing through the tail are free to leave the system and become part of the CR pool. 

From observations of Bow Shock PWNe we know that also at this stage of the evolution, the particle spectrum is very similar to what described above for their younger counterparts \citep{kaspi01,bykov03,gaensler04} (see also \cite{kargaltsev15} for a recent review). In particular, radio emission again shows a flat particle spectrum up to hundreds of GeV. In addition, numerical simulations support the idea that these low energy particles can easily escape from the system along the tail with negligible energy losses \citep{bucciantini05}. A conservative estimate of the number of pairs a pulsar releases in the ISM is then obtained by normalising the energy contained in the released pairs, described as a double power-law, to the amount of rotational energy the pulsar is left with at the time of its escape from the remnants. For typical values of the parameters this turns out to be $E_{\rm res} \approx 10^{47}$ erg. The number of leptons injected by PWNe per unit time and unit energy interval can then be estimated as:
\be
Q_{e^+}(E)\approx (2-\gamma_{pos}) {\cal R}_2\ \frac{\eta E_{\rm res}}{E_{\rm cr}^2}\left(\frac{E}{E_{\rm cr}}\right)^{-\gamma_{pos}}\ ,
\label{eq:snposinj}
\ee
where ${\cal R}_2$ is the rate of type II SN events in the Galaxy, and $\gamma_{pos}\approx 1.5$ and $E_{\rm cr}\approx$ 500 GeV describe the low-energy spectrum of pulsar produced pairs. This has to be compared with the injection of primary electrons, assumed to be accelerated by SNRs with the same spectrum as protons and in a ratio $f_{ep}$ to the latter:
\be
Q_{e^-}(E)\approx (\gamma_{el}-2) {\cal R}\ \frac{\xi_{\rm CR} f_{e,p} E_{\rm SN}}{(m_p c^2)^2}\left(\frac{E}{m_p c^2}\right)^{-\gamma_{el}}\ ,
\label{eq:snelinj}
\ee
with $\gamma_{el}\approx 2.4$, ${\cal R}$ the total rate of SNe in the Galaxy and $\xi_{\rm CR}$ the average CR acceleration efficiency of a SNR.

Since propagation affects electrons and positrons in the same way, the ratio $\chi$ will simply scale as the injection, and for energies large enough that the secondary positrons can be neglected, will read
\be
\chi\approx \frac{Q_{e^+}}{Q_{e^-}}\approx 0.035 \left(\frac{E}{30 {\rm GeV}}\right)^{1/2}
\label{eq:chiboe} 
\ee
where the last approximate equality holds for the typical values of the parameters: $\xi_{CR}=0.1$, $f_{e,p}=0.01$ and ${\cal R}_2=0.8{\cal R}$. At 30 GeV this value of $\chi$ is within a factor of 2 of the AMS-02 measurement (Fig.~\ref{fig:elposfrac}), and increases with energy, which strongly suggests that the contribution of PWNe must definitely be taken into account when wondering about the positron excess. 

A more quantitative comparison with experimental data requires taking into account the spatial distribution of PWNe and propagating the pairs using the Green function formalism \cite[]{syro59}, as explained by \cite{blasicougat}. Moreover, in order to calculate the positron ratio one needs to describe the injection and propagation of both secondary (produced by CR interactions in the ISM) and primary electrons (presumably accelerated at the same SNR shocks responsible for CR acceleration).


The rate of SN explosions is provided by \cite{faucher} ($\approx$ 2.8/century); the CR acceleration efficiency and the e/p ratio have been assumed as $\xi_{CR}=0.15$ and $f_{e,p}=0.01$, together with an injection spectrum $\propto E^{-2.4}$. One can also assume that electrons and positrons are injected by Bow Shock PWNe, formed in 80\% of the total SN explosions, after a time corresponding to the escape of the pulsar from the parent SNR, as determined by the pulsar proper velocity; the pulsar parameters that enter the normalisation of the pair spectrum (including the pulsar velocity, which determines the escape time) correspond to the best fit distributions found by \cite{faucher}; the injection spectrum can be simplistically assumed to have a slope $\gamma_{pos}=1.2$ up to an energy of 500 GeV and steepen at higher energies to $\gamma_{pos}=2.3$. Secondary electrons and positrons can be computed from modelling of the interaction with the ISM of a flux of protons as measured by AMS-02; the ISM is assumed to fill a 150 pc thick disk with a density of 1 particle ${\rm cm}^{-3}$. For all leptons the losses are computed assuming a magnetic field strength of $3~\mu$G and an energy density of IR radiation corresponding to 1 eV/${\rm cm}^{3}$.

\begin{figure}
\includegraphics[width=0.8\textwidth]{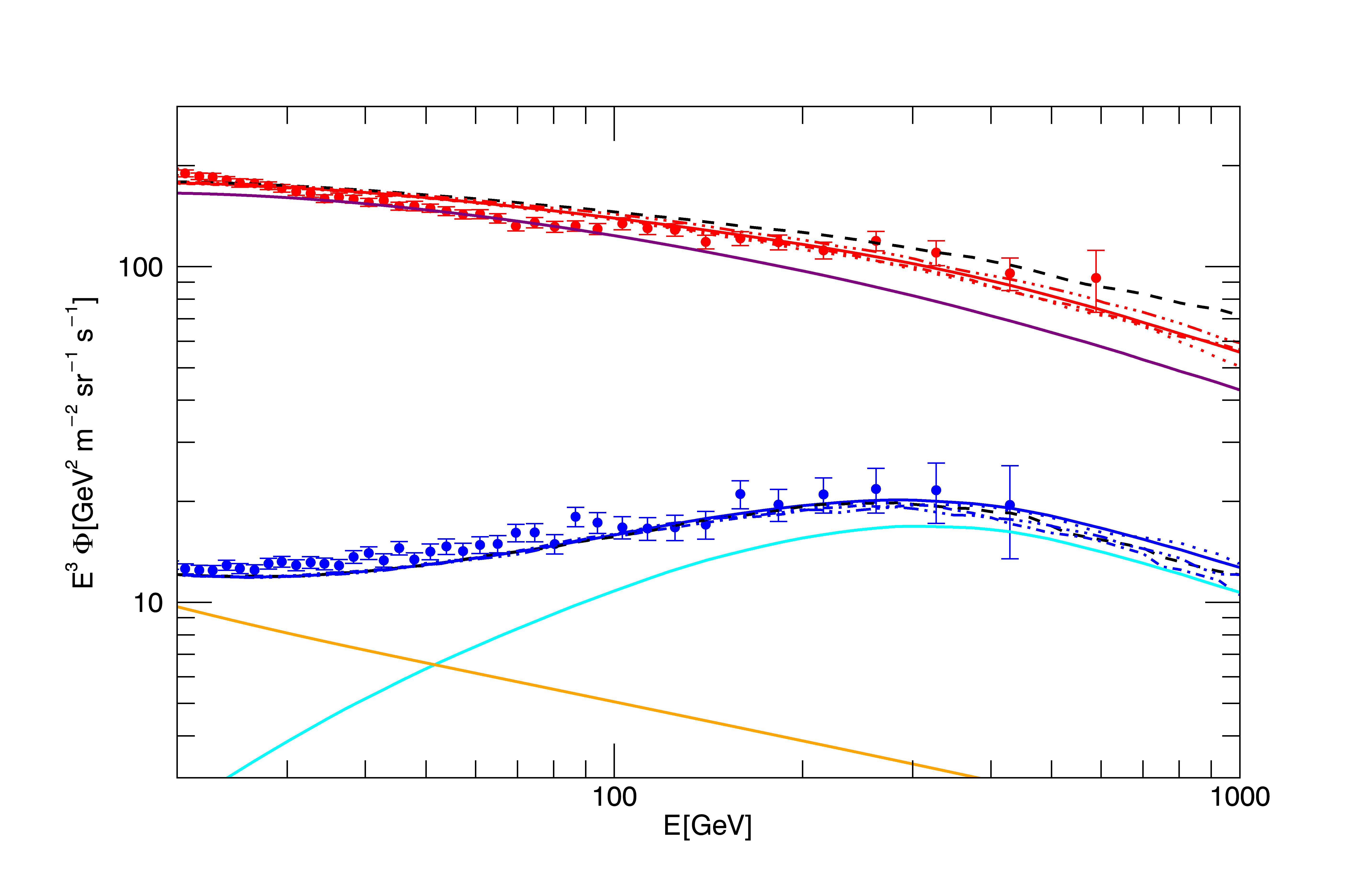}
\caption{The spectrum of electrons and positrons resulting from SNe, CR interactions in the Galaxy and release of pairs by Bow Shock PWNe is compared with the AMS-02 data. Red curves are for electrons and blue curves are for positrons. The different linetypes correspond to different realisations of the distribution of sources in the Galaxy. The thick solid red (blue) line is the spectrum of electrons (positrons) resulting from averaging over 30 different realisations. The black dashed curves represent the spectra of electrons and positrons in a particular realisation that seems to fit the data particularly well. The purple solid curve represents primary electrons only. The orange curve represents secondary electrons and positrons from CR interactions. The cyan curve represents electrons and positrons from PWNe, for an average efficiency of particle acceleration around 25\%. See text for details.}
\label{fig:elposspec}
\end{figure}

The spectra of electrons and positrons resulting from SNe, CR interactions in the Galaxy and pairs released by Bow Shock PWNe are shown in Fig.~\ref{fig:elposspec} compared with the AMS-02 data (see caption). The positron fraction calculated in this same model is plotted and compared with AMS-02 and PAMELA data in Fig.~\ref{fig:elposfrac}. It is apparent that this simple model, including the contribution of electron-positron pairs coming from pulsars in addition to standard primary and secondary electrons, can provide an excellent fit to the data. The adopted pulsar parameters are very standard and the assumed pair acceleration efficiency of PWNe (required to be of order 20-25\%) is in line with the values derived from observations.
\begin{figure}
\includegraphics[width=0.8\textwidth]{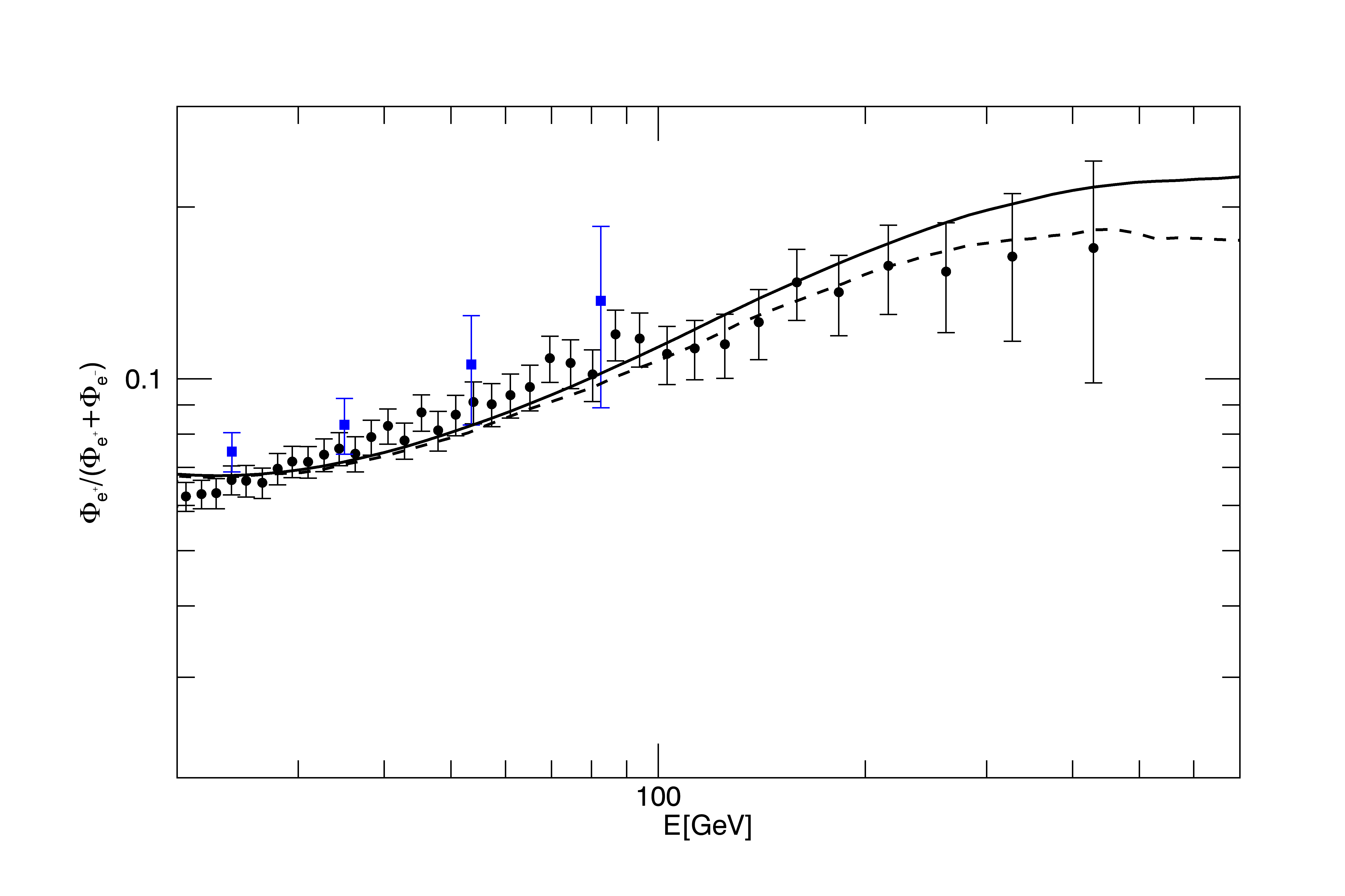}
\caption{The solid curve represents the positron fraction corresponding to the electrons and positrons spectra shown as solid red and blue curves in Fig.~\ref{fig:elposspec}.This curve results from an average over 30 different realisations of the distribution of sources in our Galaxy. The dashed line corresponds to a particular realisation that seems to go especially well through the AMS-02 data points (black circles). Also shown are PAMELA data, as blue squares.}
\label{fig:elposfrac}
\end{figure}

The simple calculation illustrated above shows in a rather straightforward manner that the contribution of pulsars to the flux of CR positrons should always be considered and addressed carefully, even while trying to assess the role of other mechanisms for the positron excess.


\section{Summary and Conclusions}
\label{sec:conclude}
 
The observational panorama in the field of CR research has never be so rich and exciting: unexpected features have been found in the spectrum of protons and helium nuclei \citep{2011Sci...332...69A,Ahn10,2015PhRvL.114q1103A}; the spectrum of He appears to be systematically harder than than of protons; the positron fraction, contrary to common wisdom, increases with energy instead of decreasing \citep{2009Natur.458..607A,2013PhRvL.110n1102A}; the $\bar p/p$ ratio is harder than expected \citep{AMS02pbar}. On the other hand, the B/C ratio measured by AMS-02 \citep{BsuC-AMS} does not show any unexpected behaviour. 

Here we tried to look at these findings in the context of non-linear theories of CR transport. As discussed several times above, this non-linearity manifests itself in two ways: first, by wave production that affects the scattering properties of the medium; and second, by dynamical action leading to motion of the ambient plasma induced by CR pressure gradients. 

The self-generation of waves is a process that is entangled in the very foundations of the theory of CR transport: particles that drift at super-Alfv\'enic speed lose some momentum to the production of waves that in turn scatter them and slow down their current. The theory describing this phenomenon allows direct connection between the micro-physics of particle scattering and the macro-physics of CR transport, whose signatures we measure at Earth.

The production of waves by CR streaming is most effective wherever the CR density and density gradient are the largest. It is expected to be at work in the acceleration region, during transport in the Galaxy and close to CR sources. In this paper we discussed separately the case of transport in the Galaxy and close to sources, while other review papers considered more in depth the case of acceleration \citep{2013A&ARv..21...70B,2014IJMPD..2330013A}. 

The spectral features observed in the H and He spectra are most likely due to transport, namely the effective diffusion coefficient felt by CRs has a different functional form at different energies. This may be due to a spatial dependence of the diffusion coefficient, as discussed by \cite{2012ApJ...752L..13T}, or to the existence of both self-generated waves and pre-existing waves, as discussed by \cite{2012PhRvL.109f1101B,2013JCAP...07..001A,2015A&A...583A..95A}. Signatures of these processes might appear in the secondary to primary ratios, such as B/C and $\bar p/p$. 

The rise of the positron fraction and the hardness of the $\bar p/p$ ratio are the two most unexpected features revealed by recent data. Some attempts at interpreting these findings as the direct signature of some peculiarity in the particle transport led to the proposal of radically new scenarios of CR propagation \citep{katz10,katz13,lipari,2010PhRvD..82b3009C,2014ApJ...786..124C,2016ApJ...827..119C}. Here we discussed these proposals and the role that non-linear effects might play in providing a physical motivation for some of the assumptions that they are based on. None of these scenarios is problem free but we believe that these are important attempts at keeping an open mind on parts of the puzzle of the CR origin that are some times taken for granted and fully understood, while there are no observations available to constrain them directly.

On the other hand, one should keep in mind that the main objections to the standard model of CR origin are based on observations that contrast with the model predictions only in its most {\it naive} version. The positron excess is easily accounted for in terms of positrons (and electrons) liberated in the ISM by pulsars that abandoned their parent supernova remnant. Not only this contribution can explain the rise in the positron fraction with energy, but it is unavoidable at some level. At the same time, the hardness of the flux of antiprotons has not been shown to be problematic in a conclusive way. In fact \cite{2015JCAP...12..039E,2015JCAP...09..023G} find that present data are still marginally compatible with the standard model once the uncertainties in the cross section for $\bar p$ production and in CR transport are taken into account. In the context of non- linear theories of CR transport the problem becomes even less severe because of the fact that diffusion is expected to acquire a weaker energy dependence at energies above $\sim 300$ GeV (relevant for the production of $\bar p$ with energy $\gtrsim 30$ GeV). These models also provide a reasonable fit to the recent data on the B/C ratio, which does not seem to be the case for most of the alternative scenarios proposed in the literature.

In conclusion, at the time being, the latest observations seem to suggest that the standard scenario of cosmic ray propagation throughout the Galaxy should be enriched in terms of new pieces of physics, rather than being abandoned in favour of radically different alternatives. Observations to be carried out in the next few years should clarify whether this is the case. 

\section*{Acknowledgements}
The authors are grateful to P. Lipari for an in-depth discussion of alternative models of CR transport and relevant cross sections for processes of relevance for CR propagation in the Galaxy. The authors are also grateful to R. Aloisio, C. Evoli, G. Morlino, S. Recchia and P. Serpico for useful discussions and continuous collaboration on these and related research topics. E.A. also acknowledges the hospitality of ISSI (Bern), by which she had the opportunity of digging into these and related subjects within teams of international experts.

\end{document}